\documentclass[a4paper,fleqn,usenatbib]{mnras} 

\usepackage{newtxtext,newtxmath}

\usepackage{lmodern} 
\usepackage{ae,aecompl}

\usepackage{graphicx}	
\usepackage{amsmath}	
\usepackage{amssymb}	
\usepackage{graphics}
\usepackage{color}
\usepackage{textcomp}     

\usepackage{natbib}
\usepackage{pdflscape}
\usepackage[export]{adjustbox}

\usepackage{hyperref}
\usepackage{rotating}
\usepackage{multicol}
\usepackage{dblfloatfix,afterpage}    

\usepackage[final]{changes}

\makeatletter
\@namedef{Changes@AuthorColor}{black}
\colorlet{Changes@Color}{black}
\makeatother

\newcommand{\etal}{et~al.\ }

\newcommand{\Ha}{H$\alpha$ }
\newcommand{\Hans}{H$\alpha$} 

\newcommand{\hi}{H{\sc i}}

\def\msun{M$_\odot$}
\def\lsun{L$_\odot$}
\def\mhi{M$_\textrm{\hi}$}
\def\mH2{M$_\textrm{\H2}$}
\def\mHc{M$_\textrm{\H2,c}$}

\def\mHZ{M$_\textrm{\H2,acc}$}
\def\mHb{M$_\textrm{\H2,bol}$}
\def\mgas{M$_\textrm{gas}$}
\def\mdust{M$_\textrm{dust}$}
\def\tdep{t$_\textrm{dep}$}
\def\H2{H$_2$}
\def\XCO{X$_\textrm{CO}$}
\def\L100{L$_{100}$}
\def\Lo160{L$_{160}$}
\def\L250{L$_{250}$}
\def\L350{L$_{350}$}
\def\L500{L$_{500}$}
\def\um{$\mu$m}
\def\Mst{M$_*$}
\def\Mism{M$_\textrm{ISM}$}

\title[Systematics in galaxy cold gas mass calibrations]{
Lurking systematics in predicting galaxy cold gas masses using dust
luminosities and star formation rates
}

\author[Janowiecki et al.]{
Steven~Janowiecki,$^{1}$\thanks{E-mail: steven.janowiecki@icrar.org (SJ)}
Luca~Cortese,$^{1}$
Barbara~Catinella,$^{1}$
Adelle~J.~Goodwin$^{2}$
\\
\scriptsize
$^{1}$International Center for Radio Astronomy Research (ICRAR), M468,
The University of Western Australia, 35 Stirling Highway,  Crawley,
WA, 6009, Australia\\
\scriptsize
$^{2}$School of Physics and Astronomy, Monash University, Clayton, Victoria, 3800, Australia
}

\date{accepted version  -- 26 Jan 2018}

\pubyear{2018}

\begin{document}
\label{firstpage}
\pagerange{\pageref{firstpage}--\pageref{lastpage}}
\maketitle

\begin{abstract}

  \noindent
We use 
galaxies from the \textit{Herschel} Reference Survey 
to evaluate commonly used indirect predictors of cold gas
masses. \added{We calibrate predictions for cold neutral atomic and
  molecular 
gas using infrared dust emission and gas depletion time methods which
are self-consistent and have $\sim$20\% accuracy (with the highest
accuracy in the prediction of total cold gas mass). 
However, modest systematic residual dependences are found in all
  calibrations which depend on the partition between molecular and
  atomic gas, and can over/under-predict gas masses by up to
  0.3~dex.  As expected, dust-based estimates are best at predicting
  the total gas mass while depletion time-based estimates are only
  able to predict the (star-forming) molecular gas mass.
Additionally, we advise caution when applying these
  predictions to high-$z$ galaxies, as significant (0.5~dex or more)
  errors can arise when incorrect assumptions are made about the
  dominant gas phase. }
Any scaling relations derived using predicted gas masses may
be more closely related to the calibrations used than to the actual
galaxies observed.



\end{abstract}




\section{Introduction}

Key to an understanding of galaxy evolution is a complete census of
the baryonic components in galaxies: the stellar populations and the
interstellar medium (ISM, gas and dust). These components are strongly
connected to each other, mainly through the process of star
formation. Atomic hydrogen gas (\hi) is the 
raw fuel for star formation, but must first cool and condense into
molecular hydrogen (\H2) before stars can form (e.g.,
\citealt{bigiel08,leroy08}, but see also \citealt{glover12}). Once formed, stars
eject gas and dust through stellar winds and also significantly
pollute the ISM with enriched material. 

The exquisite interplay between the multi-phase ISM and the star
formation cycle has been observed in nearby galaxies in great
detail. While star formation rates (SFRs) have been measured in large
samples of 
galaxies for many decades \citep[e.g.,][]{roberts63, kennicutt83,
  dr12}, more recently several
large observational programs have quantified galaxy ISM contents in a
statistical way, including 
\hi \, content \citep{catinella10,catinella13}, 
\H2 content \citep{saintonge11a}, 
and dust content \citep{dacunha10, cortese12}. These studies and others have shown many
connections between star formation and ISM phases, including that
\H2 is responsible for regulating galaxy SFRs \citep{tacconi13,
  saintonge16}, that the dust-to-gas mass ratio depends on
metallicity \citep{draine07, leroy11}, and that the dust-to-stellar
mass ratio depends on specific SFR \citep{dacunha10}.

Despite the abundance of high resolution and multi-wavelength
observations of galaxies, our understanding of the gas-dust-star cycle
is incomplete, even in the local Universe. Nonetheless, as 
observations of ISM components in galaxies at higher redshifts are
becoming increasingly feasible, we are gaining new windows into galaxy
evolution. However, the high redshift view is often limited or
partial, and requires careful ``calibration'' when making comparisons
with local galaxies.

Star formation rate is the most easily measured piece of the
gas-dust-star cycle, and samples of star-forming galaxies have been
observed out to high redshift
\citep[e.g.,][]{whitaker14,dr12}. Extensive observations have
consistently shown that the star formation history of the universe
reaches a peak at $z$$\sim$$2.5$ \citep{madau14}, but direct observations
of all ISM components are not yet as advanced
\citep[e.g.,][]{carilli13}. Observational challenges prevent the
detection of 21cm \hi \, emission in samples of galaxies beyond
$z$$\sim$$0.2$ 
\citep[][but see recent detection at $z=0.376$ from
  \citealt{fernandez16}]{highz}, but CO (a tracer of \H2) observations have
begun slowly 
reaching galaxies at higher redshifts \citep[e.g.,][]{daddi10,
  aravena12, bauermeister13, hodge13, cybulski16}.

Faced with the great difficulty (or impossibility) of directly
observing the gaseous phases of galaxy ISM at high redshift, many
observational campaigns 
have focused on \textit{indirect} estimates of ISM masses. These
estimators are based on correlations observed
between properties of local galaxies, and are then applied at higher
redshift. 

These ISM mass predictions fall broadly into two categories:
1) mass-to-light ratios using far-infrared (FIR) luminosities or
gas-to-dust ratios \citep[e.g.,][]{magdis12, eales12, scoville14}; 
and 2) starting from an integrated star formation law
\citep[e.g., the Kennicutt-Schmidt law (K-S);][]{schmidt59,kennicutt98} and
inverting the observed SFR 
(i.e., a gas depletion time) to estimate a star-forming gas mass
\citep[e.g.,][]{tacconi13, genzel15, berta16}. In the literature,
these relations are typically calibrated with samples of local
galaxies.

Recently, some groups have also explored the theoretical connections
between simulated galaxies and different ISM mass
predictions. \citet{goz17} use a fully cosmological hydrodynamical
code to reproduce the multi-wavelength emission (ultraviolet to
infrared) of galaxies, and validate some choices of indirect ISM mass
indicators. Broadly, star formation can contribute to dust heating
(making dust temperature a good tracer of SFR), but longer
wavelength dust emission is tightly connected to gas masses.

A significant limitation in the development of these indirect
relations is that each ISM mass component is usually calibrated
independently with an observable quantity, which does not allow an
analysis of possible systematic effects or residual dependencies on
other ISM properties. For example, \citet{scoville14} use \hi, CO, and
FIR observations of a sample of 12 low-$z$ spiral and starburst
galaxies in order to calibrate a predictive relationship between FIR
dust luminosity and total ISM mass (\Mism, the sum of the \hi, \H2,
and dust components), 
but their sample is too small to test for secondary
dependencies (e.g., on \mhi/\mH2 ratio, SFR, etc). By contrast,
\citet{scoville16} use the same dust 
luminosities to calibrate a prediction for molecular gas masses in a
larger 
sample of 70 galaxies. This subtle shift requires an
assumption about the relative contributions of \hi, \H2, and dust to
the total ISM mass. In a different work, the same FIR luminosities are
used to instead predict total gas mass (\mhi + \mH2) for 
36 nearby galaxies \citep{groves15}.

While economical, these indirect estimates may suffer from unknown
biases or systematic trends with other ISM properties of
galaxies. Before  
using predicted gas masses to make conclusions about the 
evolution of high-redshift galaxies, we must first carefully examine
these calibrations for any lurking systematic effects which could
introduce variations as a function of galaxy properties and mask the
true evolutionary trends.

%
%
%


We use a sample of galaxies with a \emph{complete
set of observations of all ISM components} to calibrate and test
these predictive 
relations, in order to evaluate how internally consistent these
estimators can be. We describe our local sample of galaxies and observations in
Section~\ref{sec:sample}. In Section~\ref{sec:ML} we use mass-to-light
and gas-to-dust ratios to predict ISM masses, 
and Section~\ref{sec:tdep} shows our
calibrations using depletion times. In
Section~\ref{sec:loop} we compare these two types of calibrations
with each other, and \added{in Section~\ref{sec:resid} we discuss
  residual secondary dependences on other
  quantities. Section~\ref{sec:implications} includes a  discussion of
  the scientific implications of these systematic effects, and in}
Section~\ref{sec:summary} we briefly summarize our main results.
Throughout this work we assume a $\Lambda$CDM cosmology with
H$_0$=70~km~s$^{-1}$~Mpc$^{-1}$, $\Omega$$_\textrm{M}$=0.27, and
$\Omega$$_\Lambda$=0.73.


%
%
%

\section{Sample and Data}
\label{sec:sample}

Our sample is selected from the \textit{Herschel} Reference Survey
(HRS), which was a volume-limited infrared (IR) imaging survey of 322
galaxies between 15 and 25~Mpc \citep{boselli10}. We select our
Primary Sample (PS) of N=68 galaxies to be those HRS targets with 
a complete set of observations of the ISM components to
calibrate and verify relationships between dust and gas content.
%

 Required
observations include detections in each of the \textit{Herschel} 
bands (100-500\um), detections of atomic and
molecular gas (\hi \, and CO), spectroscopic metallicity
estimates, and reliable SFRs. Further, we select only galaxies which
are not \hi-deficient. The \hi-deficiency parameter
(Def$_\textrm{\hi}$) is the logarithmic difference between the \hi\,
content expected based on a galaxy's morphological type and optical
diameter and its observed \hi\, mass \citep{haynes84}. We require our
PS sample to have Def$_\textrm{\hi}$$\le$0.5 in order to exclude
galaxies which have likely been stripped of their gas or otherwise
affected by dense environments \citep[e.g.,][]{cortese16}.
%
%
%
%
The 68 galaxies in our PS have been shown to be representative of a
volume-limited sample, and do not suffer from significant selection
effects 
\citep[see Figure~1 in][]{cortese16}.
We next briefly summarise each observed or derived quantity,
\added{their uncertainties,} and the
necessary assumptions. \added{The uncertainties on each quantity
  typically reflect the 
  errors associated with the measurement and not the calibration
  or systematic uncertainties, which may be even larger.}

\textit{Stellar mass (\Mst) and distance (D)}: \Mst \, and D 
come from \citet{cortese12}. Estimates of \Mst \, are based on the
\added{colour-dependent mass-to-light ratios of \citet{zibetti09} and
  measured from Sloan Digital Sky Survey images \citep{dr7}. 
Uncertainties on \Mst \, include only
  errors from the broadband flux measurements and not
  uncertainties from distances or mass-to-light ratios.}

\textit{Star formation rate (SFR):} We use SFRs from
\citet{boselli15}, who compiled and 
calibrated observations of star formation in the HRS. Specifically we
use their ``SFR$_\textrm{MED}$'' which combines all available SFRs
(including those from \Hans, FUV, 24\um, and radio
indicators). \added{When available, uncertainties come from the
  errors on the \Ha fluxes or radio continuum fluxes; otherwise we
  adopt a 15\% error as suggested in \S5.1.1 of \citet{boselli15}.}

\begin{figure}
  \centering
\includegraphics[width=0.90\columnwidth]{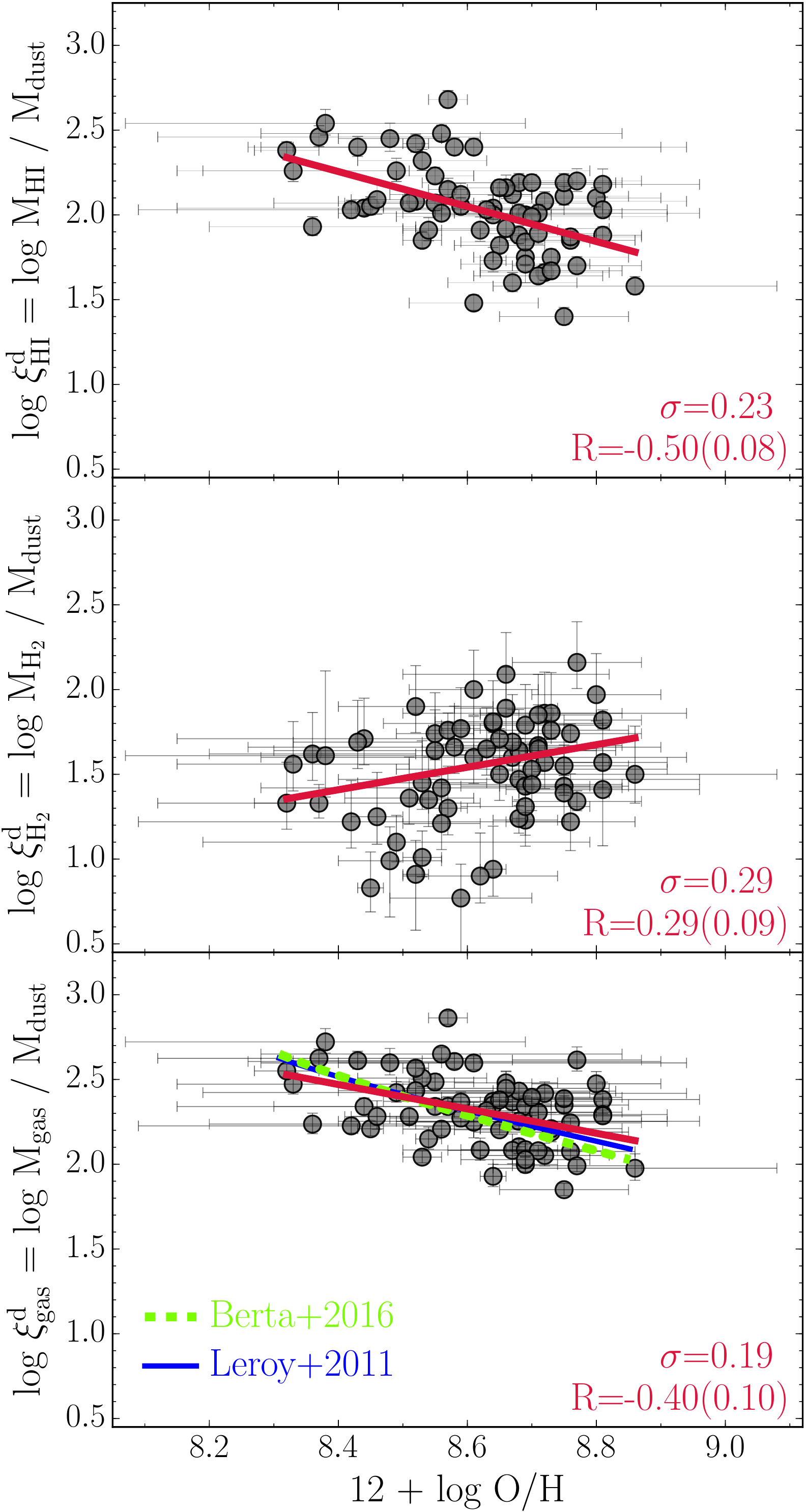}  
  \caption{\added{Calibration of $\xi$ for each gas phase as a
      function of metallicity.} Top: \mhi; middle:
    \mH2; bottom: \mgas. Our best-fit 
    relations are shown in red. The relations from \citet{leroy11} and
    \citet{berta16} are shown in blue and green, respectively,
    \added{on the bottom panel}.
  Best fit lines, 
    Pearson's correlation coefficient (R) \added{and its uncertainty,}
    and the  standard deviations 
    ($\sigma$, in dex) of points about the best-fit line are shown in
    red on each plot. 
    \label{fig:gdcal}
  }
\end{figure}

\textit{Metallicity (12+log(O/H)):} Spectroscopic gas-phase
abundances come from 
\citet{hughes13}, who observed HRS galaxies with drift-scan optical
spectra and derived oxygen abundances using the calibrations of
\citet{ke08} to be consistent with the O3N2 scale of \citet{pp04}. When
necessary, all metallicities used in this work (e.g., from
\citet{leroy11,berta16} in Figure~\ref{fig:gdcal}) are converted to
the O3N2 scale \citep{pp04} via calibrations from
\citet{ke08}. \added{Uncertainties are given by \citet{hughes13} and
  fully include the errors in converting to a common calibration of
  metallicity \citep[see \S3 of][]{hughes13}.}

\textit{Monochromatic infrared luminosities:} HRS IR data include imaging
from the Photoconductor Array Camera and Spectrometer (PACS) and the
Spectral and Photometric Imaging Receiver (SPIRE) instruments. PACS
provided images at 100\um \, and 160\um, and SPIRE at
250\um, 350\um, and 500\um. Photometry based on these images was
presented in \citet{cortese14} from PACS and in 
\citet{ciesla12} from SPIRE. We convert these flux measurements to
luminosities in solar units (\lsun). \added{Uncertainties on the
luminosities are propagated from published flux errors in
\citet{ciesla12} and \citet{cortese14}.}

\textit{Dust mass (\mdust):} We use the dust mass estimates (in \msun)
from \citet{ciesla14}. They fit spectral energy distributions (SEDs)
between 8-500\um \, using the dust models of
\citet{dl07}, \added{and provide uncertainties on their mass
  estimates.}

\textit{\hi \, gas mass (\mhi):} The atomic gas mass estimates come from
\citet{boselli14}, who compiled 21cm observations from the literature
for HRS galaxies. \added{When available, we propagate 21cm flux errors
  into \mhi \, uncertainties, otherwise we adopt the typical
  uncertainty of 15\% errors on the mass \citep[see \S6 of
  ][]{boselli14}. }

\textit{Total gas mass (\mgas):} Throughout this work we use the
following definition of total gas mass:
\mgas~=~1.36~$\times$~(\mhi~+~\mH2). This includes a 36\% correction
for helium in the total gas 
mass (\textit{not} included in \mH2 \added{alone). Uncertainties are
  propagated from \mhi \, and \mH2.}

\subsection{Molecular hydrogen masses (\mH2)}
\label{sec:h2}

\added{Estimates of molecular hydrogen gas mass come from the CO(1-0)
  observations and compilation of archival data in
  \citet{boselli14}. They use a variable conversion factor between
  L$_\textrm{CO}$ and \mH2 (\XCO) %
  which depends on the
  near-IR luminosity (measured in the H filter) of each galaxy,
  following the calibration of 
  \citet{boselli02}, not including any contribution or correction from 
  helium. Uncertainties on the mass are propagated from 
  flux errors given in \citet{boselli14}.}

\added{The variable \XCO \, from \citet{boselli02} relies
  on a number of assumptions. While a metallicity dependence
  is expected \citep{leroy09}, spectroscopic abundances were not
  available for all of the galaxies in the calibration
  sample. Instead, they demonstrated that a luminosity-dependent
  expression for \XCO \, works well for galaxies without
  metallicity measurements.
%
Their total molecular gas masses also rely
  on the assumption that molecular clouds are virialized systems
  (see Section~4 of \citealt{bolatto13} or \citealt{mckee07} for
  further details).}
%
%

\added{As a simple way to consider the range of effects of different
  conversion methods, we include three alternative prescriptions for
  \XCO \, and briefly explore their effects on gas mass
  predictions. The first,
  \mHc \, is determined using the constant conversion factor 
  from Galactic studies
  \citep[\XCO=2.3$\times$10$^{20}$~cm$^{-2}$/(K~km~s$^{-1}$),][]{strong88}. Second,
  we use the recent prescription of
  \citet{accurso17} which depends on metallicity and $sSFR$ to compute
  \mHZ. Third, \mHb \, comes from \citet{bolatto13} and depends on
  both metallicity and stellar surface density (determined from
  optical effective radius). 
   While a full discussion of the relative advantages and
  disadvantages is outside the scope
  of this work, we re-derive our main calibrations using these
  alternatives as discussed in Section~\ref{sec:alt2}. Although the
  exact values of 
  the relationships between dust luminosity and molecular gas mass
  change, these alternative prescriptions do not change the underlying
  systematic trends which are implicit in these relationships.
}

\subsection{Gas-to-dust ratio ($\xi$) for each phase}
The ratios of gas-to-dust mass are
computed using each phase (\hi, \H2, total). As this ratio 
var\added{ies} with metallicity \citep{draine07, leroy11}, we
\added{explore} the relationship between $\xi$
(\mdust/\mgas$_\textrm{(phase)}$) and 12+log(O/H) 
for each phase. Figure~\ref{fig:gdcal} shows these relationships,
\added{with uncertainties shown as error bars. Note that the
  uncertainties on the metallicity are large, as they include
  both statistical and systematic errors associated with converting
  many observations to a common metallicity scale. }

\added{We estimate the uncertainty on Pearson's $R$ correlation
  coefficient, given in parentheses on Figure~\ref{fig:gdcal} and all
  subsequent, as the
  standard deviation of the $R$ values of 10,000 random samples drawn
  from our data points. We allow for repeated values and require each
  randomly drawn sample to have the same total number of values as our
  real sample. As is the case
  throughout this work, red lines in Figure~\ref{fig:gdcal} show our
  best fits, which are found through least squares minimization of
  differences in the ordinate weighted by their uncertainties. These
  best-fitting relations are given
  below:}
\begin{equation*}
\begin{array}{l}
\log \xi^\textrm{d}_\textrm{\hi}$$=$$ 
\log \frac{\textrm{M}_\textrm{\hi}}{\textrm{M}_\textrm{dust}}$$=$$
(10.92\pm2.00)$$-$$(1.03\pm0.23)$$\times$$\textrm{(12+logO/H)}\\
\log \xi^\textrm{d}_\textrm{\H2}$$=$$ 
\log \frac{\textrm{M}_\textrm{\H2}}{\textrm{M}_\textrm{dust}}$$=$$
($-$4.17\pm2.23)$$+$$(0.67\pm0.26)$$\times$$\textrm{(12+logO/H)}\\
\log \xi^\textrm{d}_\textrm{tot}$$=$$ 
\log \frac{\textrm{M}_\textrm{tot}}{\textrm{M}_\textrm{dust}}$$=$$
(\,8.51\pm1.65)$$-$$(0.72\pm0.19)$$\times$$\textrm{(12+logO/H)}.\\
\end{array}
\end{equation*}
\noindent
%
Note that \added{the relationship between
  $\xi^\textrm{d}_\textrm{tot}$ and \textrm{12+logO/H} is the tightest 
(smallest 
scatter, $\sigma$, about the best-fit line) and also agrees
with the relations of \citet{leroy11} and \citet{berta16}. The
relationship with $\xi^\textrm{d}_\textrm{\hi}$ is the strongest
(greatest absolute value of Pearson's correlation coefficient, $R$),
owing to the 
fact that our galaxies are \hi-dominated (like most in
the local universe). The
relationship is very poor for \mH2, where the 
correlation is actually the inverse of physical expectations
\citep[e.g.,][]{leroy11} and the scatter becomes even larger.
}

\begin{figure*}
\includegraphics[angle=90,height=0.93\textheight,valign=t]{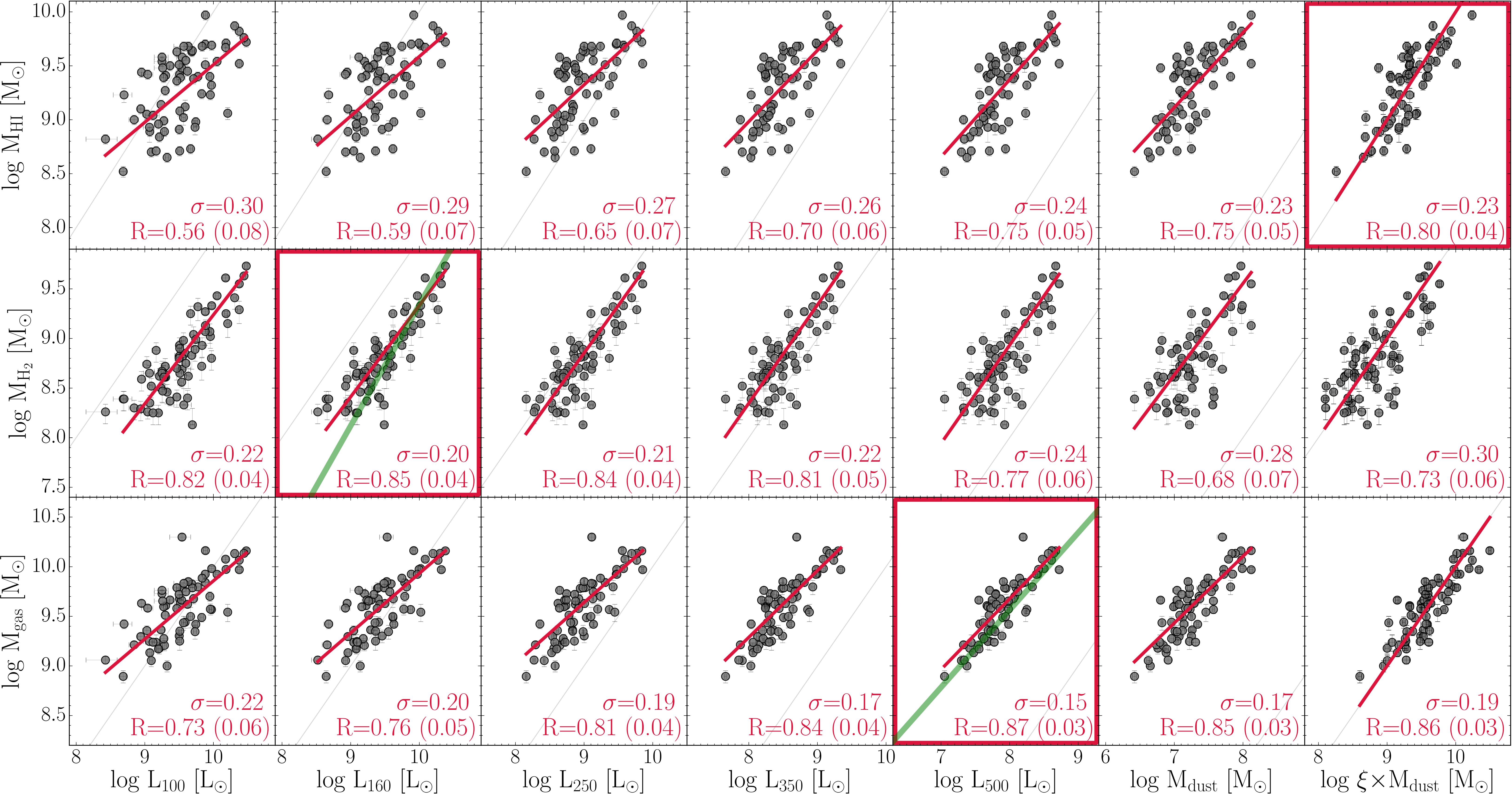}
  \caption{Correlations between gas phase masses (\hi \, in top row,
    \H2 in middle row, total \mgas \, in bottom row) and monochromatic IR
    luminosities (from 100$\mu$m to 500$\mu$m) and dust
    masses (with and without metallicity-dependent $\xi$ 
    for that phase). 
   Light grey lines show 1:1 lines in each plot, and the axes have the
     same tick sizes in all panels.
  In the last column ($\xi$ $\times$ M$_\textrm{dust}$) the
   red line is a unity line and no fitting is performed as
  $\xi$ was already fit in Figure~\ref{fig:gdcal} \added{for each
    phase}.
    The strongest and tightest relationships 
    are indicated by bold boxes.
  Wide green lines show \added{the agreement with relations from
    \citet{groves15}}.
    \label{fig:ML}
  }
\end{figure*}

%
%


\section{Gas mass estimates from L$_\textrm{IR}$ \& \mdust}
\label{sec:ML}

We start by considering the simplest possible correlations between
cold gas masses (atomic, molecular, and total) and monochromatic IR
luminosities, as shown in the first five columns of
Figure~\ref{fig:ML}. We also consider correlations with dust mass,
with and without the \added{metallicity-dependent $\xi$, 
  as shown in the last
  two columns. 
In all cases we carry out least squares fitting weighted by the
  uncertainties from both axes. The minimization is only done for
  differences in the ordinate as these relationships are intended for
  use as predictions between ``known'' IR luminosities or dust masses
  and ``unknown'' ISM phase masses.}

\added{We caution against the blind application of these equations to
  diverse populations of galaxies. As will be shown in the following
  sections, calibrations of cold gas mass predictions are susceptible
  to a number of biases and systematic residual dependences on other
  physical properties. The best-fit predictions we derive from our
  sample may not be appropriate to use across all types of
  galaxies. Further discussions of these biases and their effects are
  included in Sections~\ref{sec:resid} and \ref{sec:implications}.}

\subsection{\mhi \, estimates from dust}

Across the M/L estimates of \mhi, 
the \added{strength and scatter improve} continuously with increasing
wavelength such that 
the best monochromatic predictor is \L500. \added{Not only does \L500
  have a value of $R$$=$$0.75\pm0.05$ which is statistically larger than
  the next best value at $R$$=$$0.70\pm0.06$, but it also has a scatter
  which is $\sim$8\% better than the next best.} This is expected, as
\L500 is most closely connected to emission from the diffuse
ISM. Multiplying
\mdust \, by the metallicity-dependent $\xi^\textrm{d}_\textrm{\hi}$ 
gives an even better prediction of \mhi\added{, both in terms of $R$
  and $\sigma$}. Our best three relationships are given below,
\added{where $\xi^\textrm{d}_\textrm{\hi}$  
 was determined in
  Section~\ref{sec:sample}.}

\noindent
\begin{equation*}
\begin{array}{l}
\log\frac{\text{\mhi}}{\text{\msun}} = 
(0.72 \pm0.09) \left( \log\frac{\text{\L500}}{\text{\lsun}} -9
\right)+ (10.09 \pm0.08) \\
\log\frac{\text{\mhi}}{\text{\msun}} = 
(0.69 \pm0.13) \left( \log\frac{\text{\mdust}}{\text{\msun}} -9
\right) + (10.50 \pm0.07) \\
\log\frac{\text{\mhi}}{\text{\msun}} = 
\log \xi^\text{d}_\text{\hi} + 
\log \left( \frac{\text{\mdust}}{\text{\msun}} \right)  \\
\end{array}
\end{equation*}
\noindent
\added{Here and in all subsequent equations we use mass and
  luminosity in solar units and use years as the denominator in all
  SFRs.}

\subsection{\mH2 estimates from dust}

Among the M/L estimates of \mH2 the 160\um \, luminosity is
best\added{, because although its $R$ value is not significantly
  different from that of L$_{100}$ or L$_{250}$, its $\sigma$ is the
  smallest. The 
  longer wavelength luminosities become increasingly poor estimates of
  \mH2}. This connection between  
molecular gas and shorter wavelengths is consistent with the use of
L$_\textrm{IR}$ as an SFR indicator \citep[e.g.,][]{reddy10}. Using
\mdust \, instead (with or without $\xi$) 
gives an even poorer predictor of \mH2.

For comparison, the relationship between \mH2 and
\Lo160 from \citet{groves15} is also shown in Figure~\ref{fig:ML} as a
wide green line, and agrees with our data points. Note that their
best-fit relation has a steeper slope than ours, as their sample
includes a number of dwarf galaxies (\Mst$<$$10^9$\msun) with small
molecular gas masses (\mH2$\sim$$10^7$\msun) which drive this
steeper slope. If these dwarf galaxies were removed to match the
stellar mass range of our sample, the \citet{groves15} best-fitting
relationship would be consistent with ours. Our best fitting
relation with \Lo160 is given below, along with the \L500 prediction
(as it is more commonly used).
\begin{equation*}
\begin{array}{l}
\log\frac{\text{\mH2}}{\text{\msun}} = 
(0.92 \pm0.04) \left( \log\frac{\text{\Lo160}}{\text{\lsun}} -9
\right)+ (8.41 \pm0.04) \\
\log\frac{\text{\mH2}}{\text{\msun}} = 
(0.99 \pm0.05) \left( \log\frac{\text{\L500}}{\text{\lsun}} -9
\right)+ (9.92 \pm0.06) \\
\end{array}
\end{equation*}

\begin{figure*} 
\includegraphics[width=0.80\textwidth]{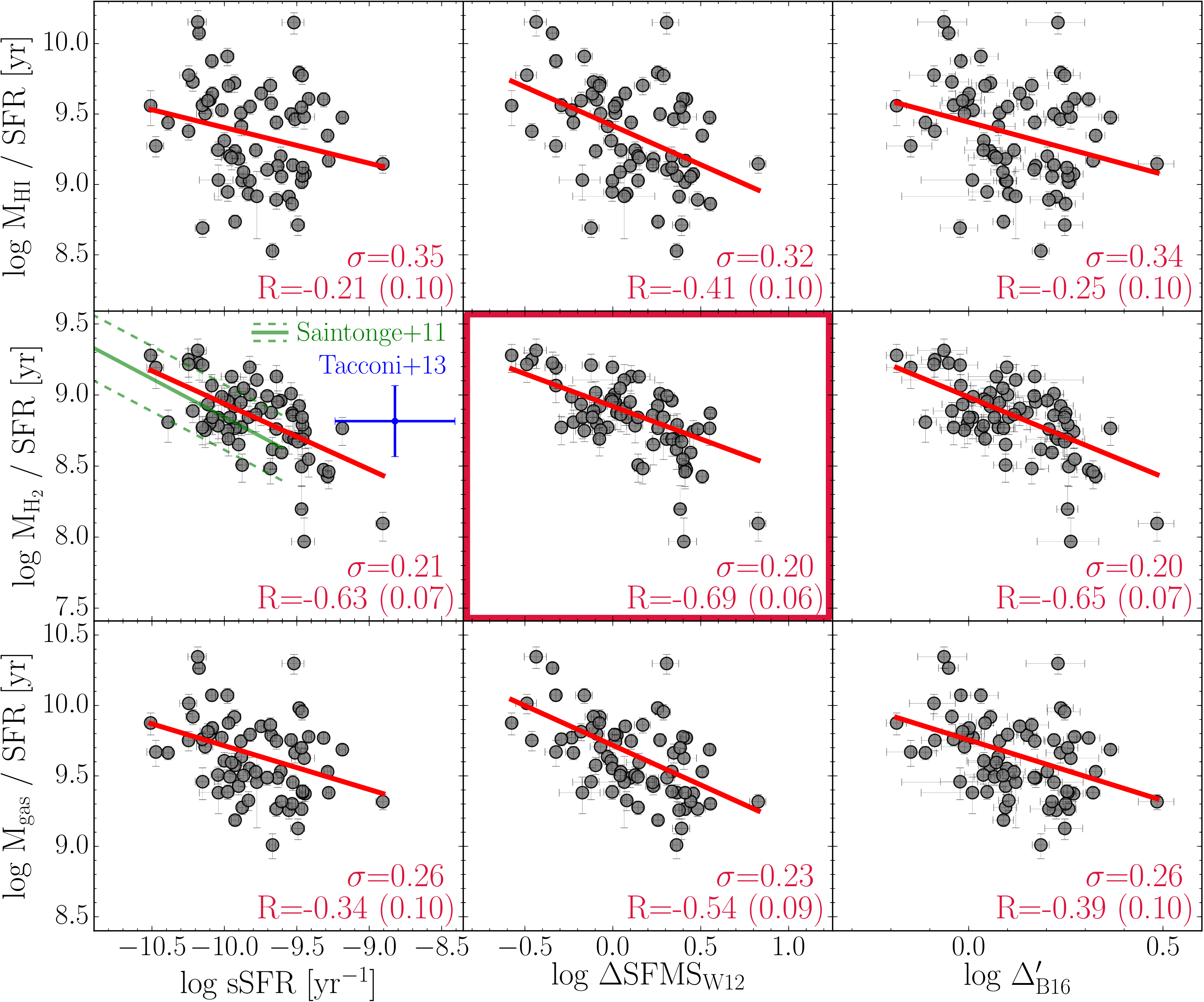}   
  \caption{Gas mass divided by SFR (i.e., depletion time) using various
        gas phases is plotted against SFR-related quantities.
    Left column x-axis shows specific SFR (sSFR), and our \mH2/SFR values
    agree with the 
      average value from \citet{tacconi13} and the trend from
      \citet{saintonge11a}.
    \added{Center column shows $\Delta$SFMS using the
    \citet{whitaker12}} SFMS.
    Right column shows 
    $\Delta^\prime_\textrm{B16}$
    from \citet{berta16}, a weighted
    combination of sSFR and \Mst.
  \label{fig:sfr}
  }
\end{figure*}

\subsection{Total \mgas \, estimates from dust}
\label{sec:mgas}

The estimates of \mgas \, show a similar behavior to those of \mhi, as
\hi \, is the dominant gas component in our galaxies. The best \mgas \, 
predictor is \L500\added{; although its $R$ value is not significantly
  better than the $\xi$-based estimates, it does have smaller 
  scatter in the relationship. This} is consistent with its connection
to diffuse 
ISM emission. The \mdust \, estimates of \mgas \, are only slightly
worse, perhaps due to the additional uncertainties from fitting SEDs to
determine dust masses. Regardless of the choice between \L500 and
\mdust, these correlations with \mgas \, are the strongest and
tightest of any combination of gas phase and indicator discussed thus
far. Our three best-fitting \mgas \, relations are given below,
\added{where $\xi^\textrm{d}_\textrm{tot}$
was determined 
  in Section~\ref{sec:sample}.}

\begin{equation*}
\begin{array}{l}
\log\frac{\text{\mgas}}{\text{\msun}} = 
(0.71 \pm0.05) \left( \log\frac{\text{\L500}}{\text{\lsun}} -9
\right)+ (10.39 \pm0.05) \\
\log\frac{\text{\mgas}}{\text{\msun}} = 
(0.67 \pm0.08) \left( \log\frac{\text{\mdust}}{\text{\msun}} -9
\right) + (10.78 \pm0.05) \\
\log\frac{\text{\mgas}}{\text{\msun}} = 
\log \xi^\text{d}_\text{tot} +
 \log \left( \frac{\text{\mdust}}{\text{\msun}} \right)\\
\end{array}
\end{equation*}

\citet{groves15} found a similar relationship between \L500 and
\mgas \, as ours, as shown by the wide green line in
Figure~\ref{fig:ML}. Their relationship agrees with our data
points after it is corrected to include helium to match our convention.

Going forward, we prefer the \L500 estimates of \mgas, as it gives the
\added{best} correlation and also does not require any
additional assumptions to determine dust mass. For \mH2 estimates, we
prefer \Lo160 if  available, but \L500 is acceptable. 
However, there are significant residual trends in these
predictions, as discussed in Section~\ref{sec:resid}.


\section{Depletion time estimates}
\label{sec:tdep}

\begin{figure*}
\noindent 
\includegraphics[width=0.98\textwidth]{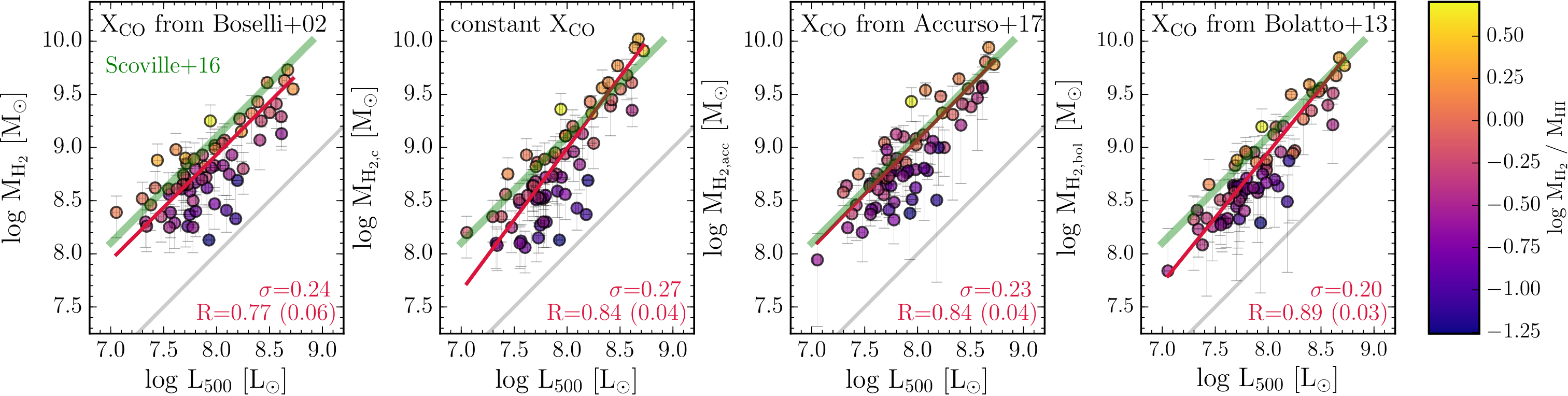}\\  
\hspace{2.5pt}      
\includegraphics[width=0.873\textwidth]{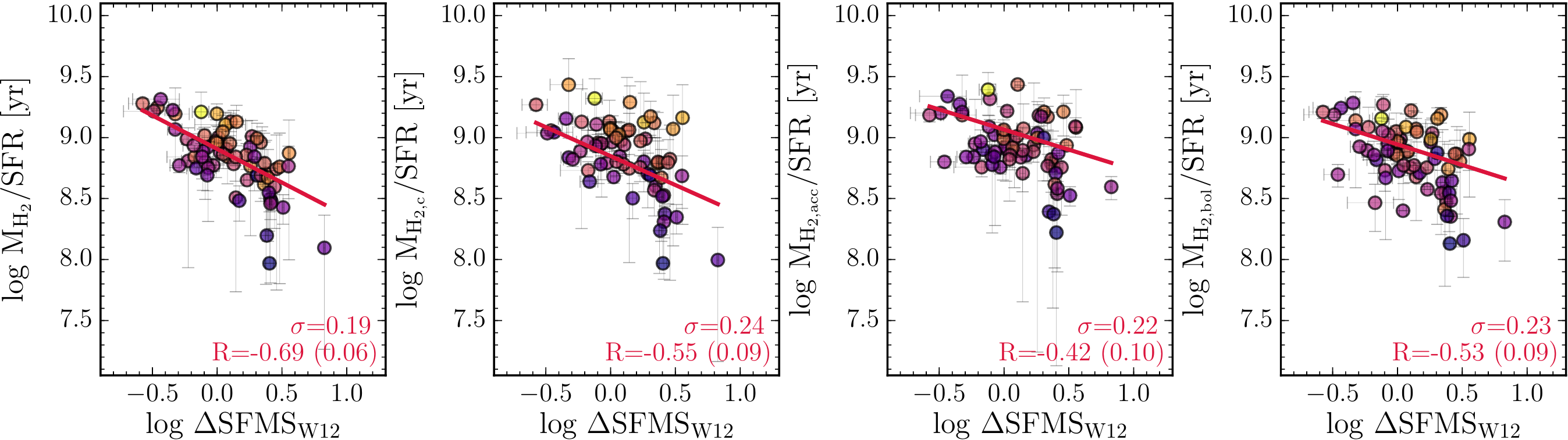} \hfill  
  \caption{
  Top row: correlations between L$_{500}$ and \mH2, using 
      different \XCO \, conversion factors. Wide green line
      indicates the relationship from \citet{scoville16}, which uses a
      constant \XCO.
  Bottom row: Correlations between depletion time (\mH2/SFR) and
     $\Delta$SFMS, using the same prescriptions.
  In all panels, points are colored by their \mH2/\mhi \, ratio, using
     the \mH2 prescription from that panel.
    \label{fig:H2c}
  }
\end{figure*}

As a second method to estimate gas mass without direct observations,
we can use the inverted integrated Kennicutt-Schmidt law (i.e., gas
depletion time, t$_\textrm{dep}$). We start by using each galaxy's SFR
as a predictor of its \added{amount} of (star-forming) gas. In order to do
this, we must determine the distance between each galaxy and the
so-called star-forming main sequence 
\citep[SFMS;][]{salim07,noeske07}. The SFMS is typically defined as the
ridge-line in the SFR-\Mst \, diagram and is populated by 
``star-forming'' galaxies. We use two different definitions of
SFMS. In general, the SFMS can be parameterized as follows.
\begin{equation*}
\footnotesize
\log \frac{\textrm{SFR}_\textrm{MS}}{\textrm{\msun}/\textrm{yr}} = 
\alpha
\left( \frac{\textrm{\Mst}}{\textrm{\msun}} - 10.5 \right)
+ \beta
\end{equation*}
Using UV+IR SFRs for $\sim$22,000 galaxies at $0<z<2.5$,
\citet{whitaker12} derived a redshift-dependent SFMS with parameters
given below.
\begin{equation*}
\alpha_\textrm{W12}(z) = 0.70-0.13z, \, \, \, \, \, \, \, \, \, \, \, \,
\beta_\textrm{W12}(z) = 0.38 + 1.14z - 0.19z^2
\end{equation*}
\noindent
We also \added{considered} the SFMS from \added{\citet[][see also
    Janowiecki \etal in prep.]{catinella18},} 
 which is based on UV+IR SFRs of
a low redshift sample of $\sim$1200 stellar-mass selected galaxies
from xGASS \citep{xenv}, 
with best-fit parameters given below.
\begin{equation*}
\alpha_\textrm{J17} = 0.656, \, \, \, \, \, \, \, \, \, \, \, \,
\beta_\textrm{J17} = 0.162
\end{equation*}
While this
is based on a representative sample of local
(z$\sim$0) galaxies, the \citet{whitaker12} SFMS includes a
redshift dependence, making it more useful in extending our
predictive relationships for ISM masses to galaxies at higher
redshifts. We \added{use only the \citet{whitaker12} SFMS going
  forward, as both give similar results for our sample}.


Using \added{this definition} of SFMS, we compute $\Delta$SFMS, which
quantifies each galaxy's distance above or below the SFMS at
its \Mst \, and redshift.
\begin{equation*}
\log\Delta\textrm{SFMS} = 
\log\frac{\textrm{SFR}}{\textrm{M$_\odot$/yr}}  
- 
\log\frac{\textrm{SFR}_\textrm{MS}\textrm{(\Mst,$z$)}}{\textrm{M$_\odot$/yr}} 
\end{equation*}


In a similar vein, \citet{berta16} derive an expression for depletion
times based primarily on $\Delta$SFMS, but also including a dependence
on \Mst \, and redshift (their Equation~3). We show this quantity as
$\Delta^\prime_\textrm{B16}$.

Figure~\ref{fig:sfr} shows our observed depletion times
(
\mgas/SFR) using different gas masses (\hi, \H2, total). These are
plotted against sSFR \added{(left column), $\Delta$SFMS (center
  column), and $\Delta^\prime_\textrm{B16}$ (right column)}. As with
Figure~\ref{fig:ML}, best-fit lines and quality estimates are shown on
each panel. 
\added{
Note that for each predictive indicator shown on the x-axis, the
relation with \mH2  has the tightest scatter and strongest
correlation. Unsurprisingly, the SFR-based indicators are less closely
related to \mhi. Since our galaxies are \hi-dominated, the
predictions for \mgas \, are similarly weak. Again we caution against
blindly applying our calibrations to populations of galaxies which are
significantly different from those in our sample. We next focus on
two of these predictions of \mH2.
}

\begin{figure*}
\centering
\includegraphics[angle=0,width=0.99\textwidth,valign=t]{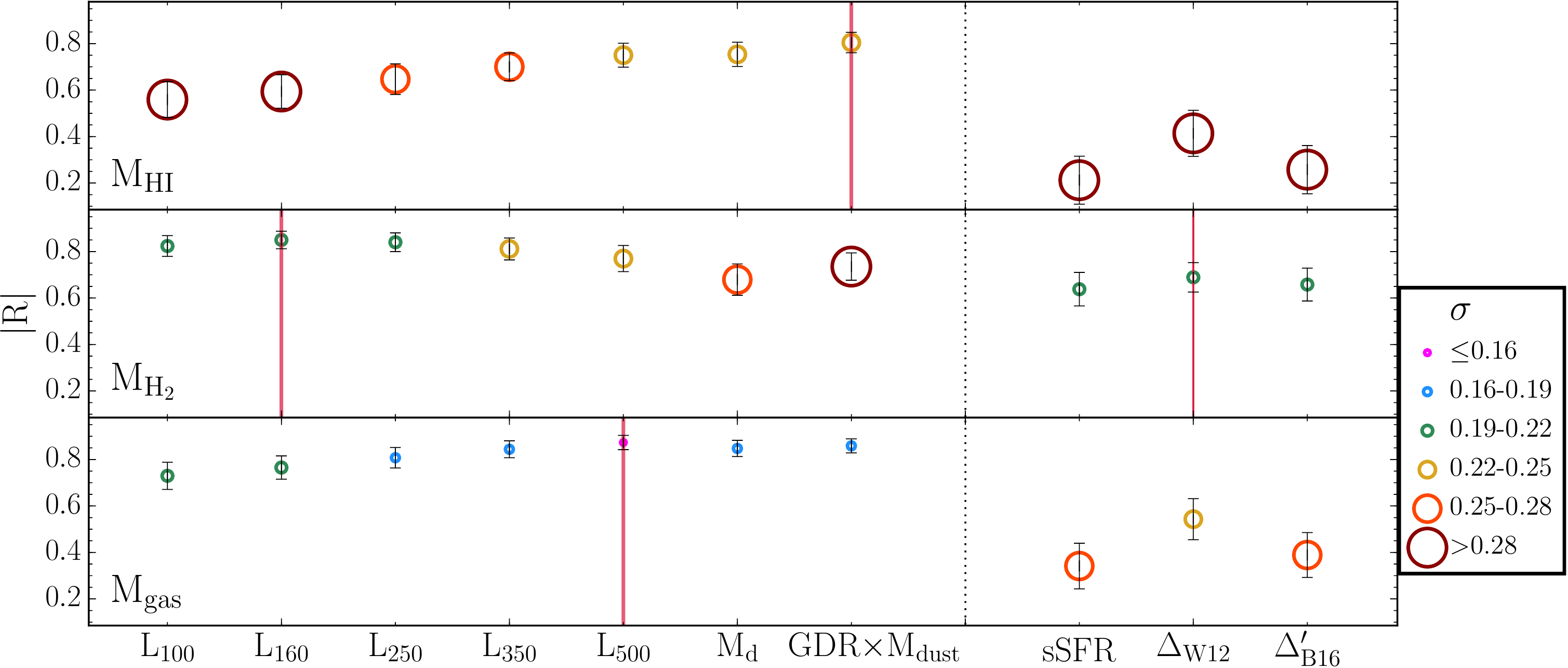}
  \caption{\added{The absolute value of Pearson's R value is shown
      for} each indicator,  
    from the correlations shown in Figures~\ref{fig:ML} and
    \ref{fig:sfr}. The size and color of each point corresponds to its
    \added{scatter 
    ($\sigma$, in dex)}, as indicated in the legend. The best
    relationships are marked with solid red  vertical
    lines. \added{Uncertainties on $|$R$|$ are shown as error bars and
      show the standard deviations from 10,000 random samplings.}
    \label{fig:trend}
  }
\end{figure*}

\subsection{\mH2 estimates from t$_\textrm{dep}$(sSFR)}
\added{
The relations shown in Figure~\ref{fig:sfr} are consistent with
previous results, as seen in 
the panel showing \mH2/SFR vs sSFR. Here the constant depletion time
estimate from \citet{tacconi13} is shown (at the average sSFR of their
sample), as well as the relation from
\citet{saintonge11a}, both of which are consistent with our
observations. Our best-fitting relationship between \mH2/SFR (in units
of years) and sSFR (in inverse years) is given below.
}
\begin{equation*}
\mathrm{log}~\frac{\textrm{\mH2}}{\textrm{SFR}} = 
 -0.46 (\pm0.74) \times
\mathrm{log}~\textrm{sSFR} 
 + 4.32 (\pm0.08)
\end{equation*}

\subsection{\mH2 estimates from t$_\textrm{dep}$($\Delta$SFMS)}

\added{The strongest and tightest correlation is found between \mH2/SFR
  and
$\Delta$SFMS,}
as the \H2 is
directly involved in current star formation. The
best-fitting relationship between $\Delta$SFMS$_\textrm{W12}$ and
\mH2/SFR (in years) is given below.
\begin{equation*}
\mathrm{log}~\frac{\textrm{\mH2}}{\textrm{SFR}} = 
 -0.46 (\pm0.03) \times
\mathrm{log}~\Delta\textrm{SFMS}_\textrm{W12} + 8.92 (\pm0.08)
\end{equation*}
\noindent
This \added{relationship has similar scatter and strength 
  as the}
\L500--\mgas \, and \mdust--\mgas \, relationships determined in
Section~\ref{sec:ML}, but also is affected by the residual trends
discussed in the following section.

\subsection{Alternative \mH2 prescriptions}
\label{sec:alt2}


\added{We briefly consider the effects of alternative \XCO \, 
conversion factors to determine \mH2 from CO
observations, as introduced in Section~\ref{sec:h2}. In addition to
the luminosity-dependent \XCO \,
we primarily use, we show here the results of using a constant
conversion factor (\mHc), 
the metallicity-dependent conversion (\mHZ) from
\citet{accurso17}, and the conversion from \citet{bolatto13} (\mHb)
which depends on metallicity and stellar surface
density (determined from optical effective radius).
Figure~\ref{fig:H2c} shows the predictive relations 
for these determinations of \mH2 using L$_{500}$ (top row) and
depletion time from $\Delta$SFMS (bottom row). In all cases, the \mH2
values do not include the contribution from helium.}

\added{Using different conversion factors does not dramatically
alter these predictive relationships. Adopting the
metallicity-dependent conversion from \citet{accurso17} moderately
improves the strength (and scatter) of the L$_{500}$ prediction, but
it significantly worsens the strength of the relationship with
depletion time. Similarly, the \citet{bolatto13} prescription
marginally improves the strength and scatter of the \L500 relation and
worsens the trend with $\Delta$SFMS. }

\begin{figure*}  
\includegraphics[width=0.99\textwidth]{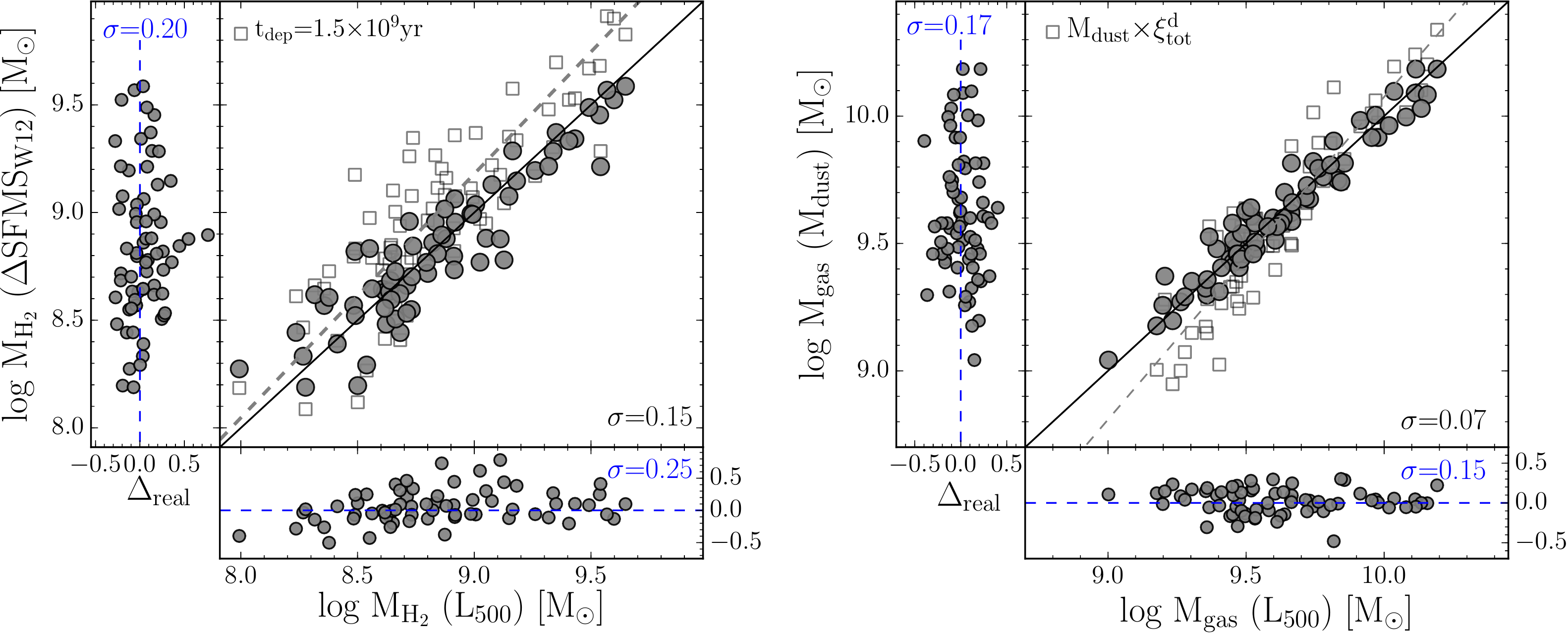}\\
  \caption{\mH2 (left) and \mgas \, (right) comparisons from our
    calibrations. In both main panels, our predictions for the gas
    phase masses are shown, using different indicators on both
    axes \added{($\sigma$ shows the scatter about the unity
      line, in dex)}. The residual panels show the differences between the 
    predicted and measured gas phase masses ($\Delta_\textrm{real}$).
\added{  Also shown 
(in light grey boxes, and best-fitting dashed lines) are
     the results when using alternative calibrations for the y-axis
     quantities. In the left panel, small boxes show the effect of 
    assuming a constant depletion time. 
    In the right panel, small boxes
     show the results of using $\xi^\textrm{d}_\textrm{tot}$ which depends on
     metallicity (and assuming \mgas=\mdust $\times$$\xi$).  }
    Note that the two estimates of \mgas \, (from \mdust \, and from
    \L500) are better correlated with each other
    ($\sigma$=0.07) than they are with \mgas, as \L500 also tracks
    the dust content.
    \label{fig:loop}
  }
\end{figure*}

\added{
Also shown in the top row of Figure~\ref{fig:H2c} is the \L500-based
\mH2 prediction from \citet{scoville16}, which is determined from
observations of 70 galaxies between $z=0-2$. That calibration
adopts the constant Galactic value of \XCO, so it is most
appropriate to compare it with the left middle panel. As expected,
the \citet{scoville16} relationship works better for galaxies with
\mH2$\gtrsim$\mhi; 
 \hi-dominated galaxies can fall as 
much as an order of magnitude below the prediction. This illustrates 
the difficulty in using FIR dust luminosity to predict \mH2 alone --
without knowledge of the atomic-to-molecular ratio, it is difficult to
estimate the mass in each phase from a dust luminosity. }



\section{Closing the loop}
\label{sec:loop}

Figure~\ref{fig:trend} visually summarizes \added{the quality of} the
predictive 
relationships calibrated thus far. Above each indicator on the x-axis
(i.e., monochromatic luminosity, dust mass, sSFR, or depletion time
estimate) the \added{strength and scatter} of its correlation with
each gas mass \added{are} shown 
(for \hi, \H2, and total gas). The y-axis position
shows the \added{absolute value of Pearson's R value, and the size of
  each point corresponds to the standard deviation of our observations
  about the best-fit (or unity) line ($\sigma$). Larger
symbols show relationships with higher scatter.}

As is shown in Figure~\ref{fig:trend} (and in Sections
\ref{sec:ML} \& \ref{sec:tdep}), \mhi \, is better predicted by longer
wavelength IR emission (reaching the best prediction at 500\um), and
slightly better still by dust mass and metallicity. As atomic gas is
less directly connected to ongoing star formation, predictions based
on depletion time are poorer for \mhi.  Molecular gas mass is best
predicted either by \Lo160 
or depletion time estimates, both of which have similar scatter but
the M/L method gives a stronger correlation, partly due to its larger
dynamic range. Total gas is best predicted by \L500, and depletion
time estimates perform poorly, as the galaxies in our sample are
\hi-dominated.

We next compare the two methods and evaluate whether these separate
predictions are mutually consistent. Figure~\ref{fig:loop} shows two
different predictions of \mH2 (left panel, from $\Delta$SFMS and
\L500) and two different predictions of \mgas \, (right panel,
\added{from} \mdust \, and \L500). Since all of these estimates 
have been calibrated using the same sample of galaxies, their good
consistency is not surprising.

The two estimates of \mH2 are observationally independent, as the
t$_\textrm{dep}$ prediction depends only on SFR and \Mst, which are
separate from the observations of \L500. It is possible to calibrate
those two predictive relationships so that mutually consistent
estimates of \mH2 
are obtained, whether using SFR or FIR luminosity. 
\added{Note that using a
fixed depletion time \citep[e.g.,][]{tacconi13} gives a relationship
which is $\sim$15\% steeper than the 1:1 line, and which over-estimates
\mH2(\tdep) at higher \H2 masses.}

The two estimates of total \mgas \, are less independent, as the \L500
emission is closely related to the dust content. As such, the two
indicators correlate with each other ($\sigma$=$0.07$~dex) more
tightly than with \mgas \, ($\sigma$=$0.15$-$0.17$~dex). This is
reassuring as it means the choice of indicator is not 
crucial, and both \L500 and \mdust \, are good estimators of
\mgas. 
%
Note also that adopting a metallicity-dependent
$\xi^\textrm{d}_\textrm{tot}$ (as derived in Section~\ref{sec:sample})
introduces a $\sim$25\% steeper slope to the relation between 
\mgas(\mdust) and \mgas(\L500). This systematic difference
\added{implies that $\xi^\textrm{d}_\textrm{tot}$ scales with galaxy
  stellar mass (via the mass-metallicity relation), as has   been 
  shown by \citet{cortese16} and \citet{remy13}}. In order to
eliminate the slope difference between these two estimates, the
metallicity-dependent $\xi^\textrm{d}_\textrm{tot}$
relationship would need a 3 times steeper slope,
which dramatically increases the scatter of the points around the
unity line in Figure~\ref{fig:loop}.

These comparisons demonstrate that \added{the choice of} calibration method
can have significant effects on the indirect predictions of cold gas
masses. Even within the same sample of galaxies there can be
systematic deviations when using different assumptions, such as a constant
depletion time or a metallicity-dependent $\xi$. Nonetheless, with
appropriate choices, independent predictive methods can be calibrated
to produce estimates which are in agreement with each other.


\section{Residual trends in calibrations} %
\label{sec:resid}

These predictive relationships perform as expected in our sample of
local, star-forming, \hi-dominated galaxies. However, we are most
interested in applying these relationships to galaxies at higher
redshifts, where the \added{partition between atomic and molecular gas
  is unknown and other physical properties may differ (e.g., \Mst,
  sSFR).} 
%
%
%
\added{
  Across our sample, the ratio of \mH2/\mhi \,
  varies between $\sim$3\% and $\sim$300\%,
which is similar to the range of values observed in galaxies from the
xGASS sample (Catinella
\etal submitted). Galaxies in xGASS also show a weakly increasing
 median value of the
molecular-to-atomic ratio as a function of stellar mass, from $\sim$10\%
at $10^9$\msun \, to $\sim$30\% at $10^{11.5}$\msun.}



\subsection{Quantifying residual systematics}

\added{Quantifying the strength of any residual secondary
  dependences requires careful 
  parametrization to avoid being affected by underlying
  dependences between \mgas \, and its constituent phases. For
  example, plotting the accuracy of the \L500 prediction of \mhi \, as
  a function of the \mH2/\mhi \, ratio has a strong intrinsic
  correlation (from the inclusion of \mhi \, on both axes) which
  makes it difficult to directly interpret the residuals in a physical
  sense. 
  We explore these intricacies in a
  Monte Carlo analysis described in Appendix~\ref{sec:app}, and
  demonstrate that our approach successfully quantifies physical
  residual dependences without suffering from these effects.
}

\begin{figure*}
\centering
\includegraphics[height=8.5cm]{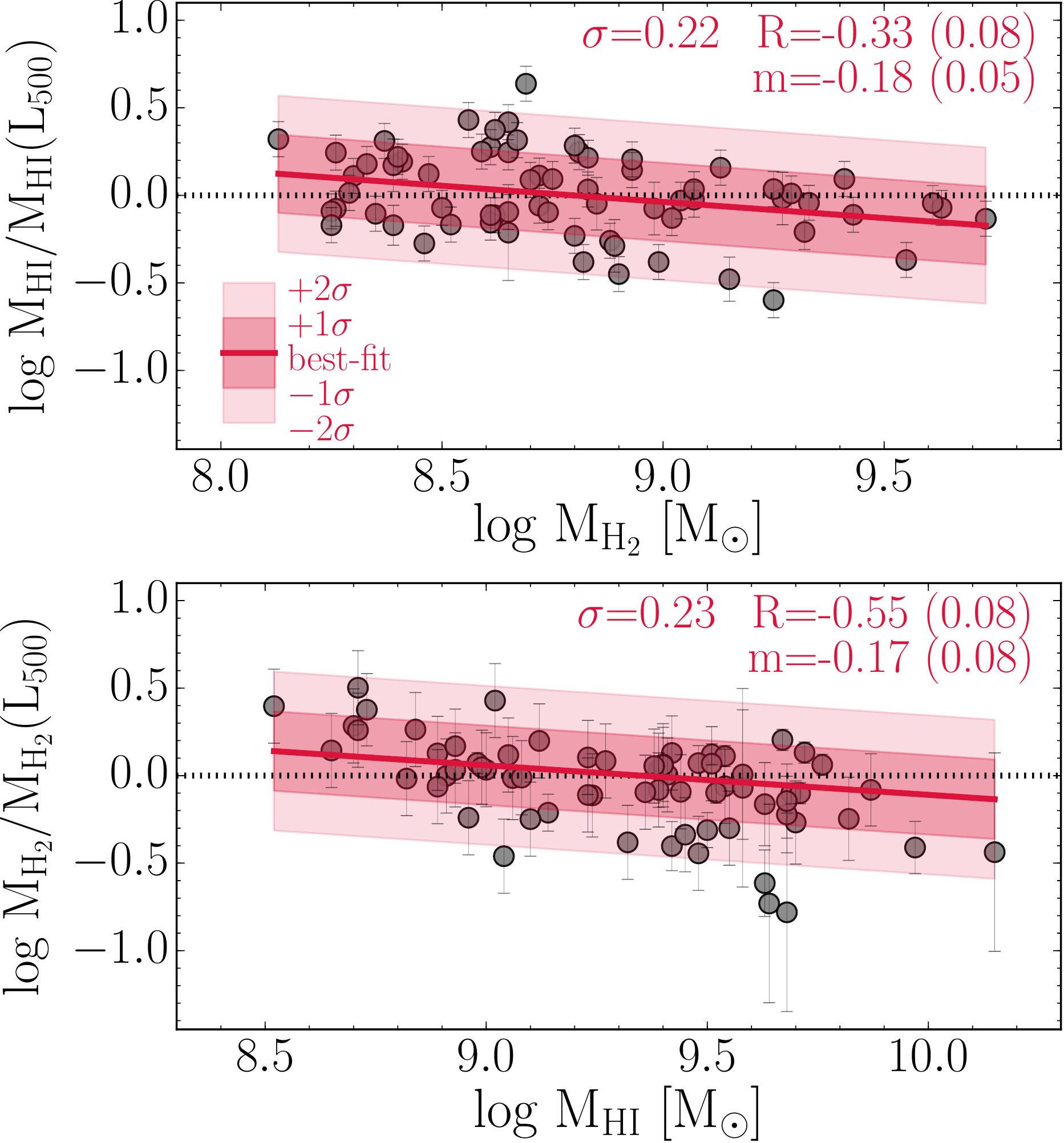} \hspace{15pt}  
\includegraphics[height=8.5cm]{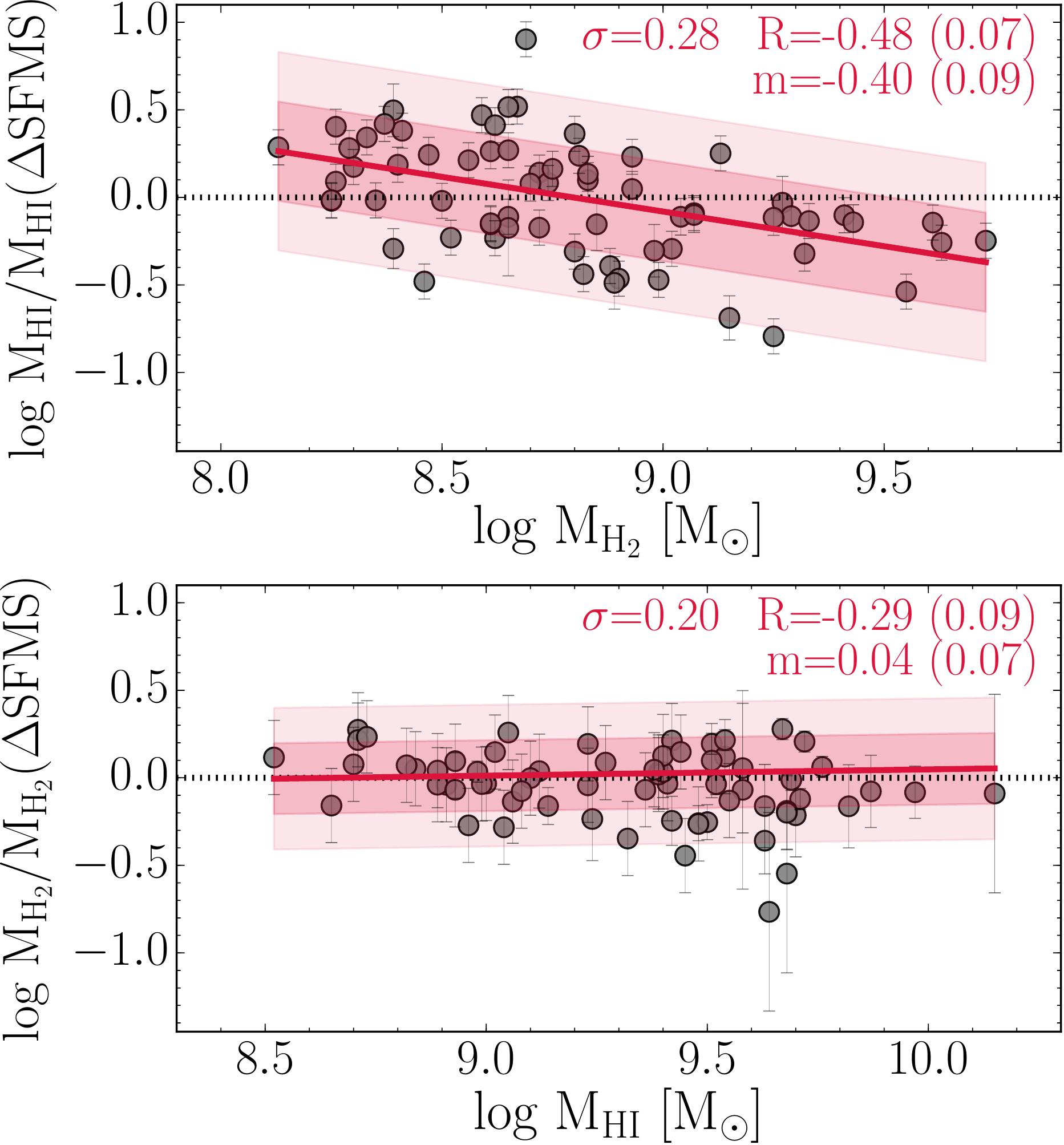}\\      
\vspace{5pt}
\includegraphics[width=7.75cm]{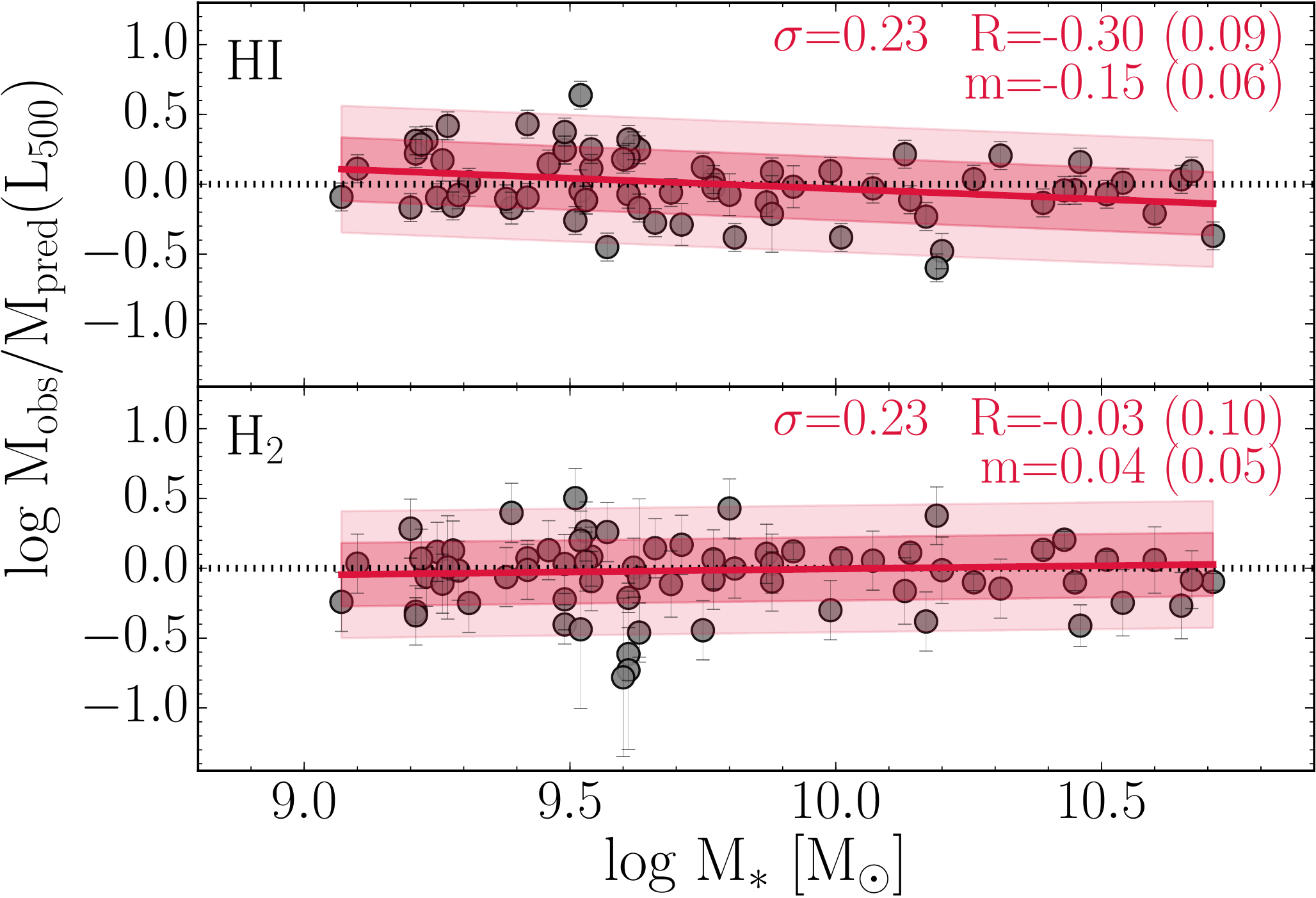} \hspace{15pt}  
\includegraphics[width=7.75cm]{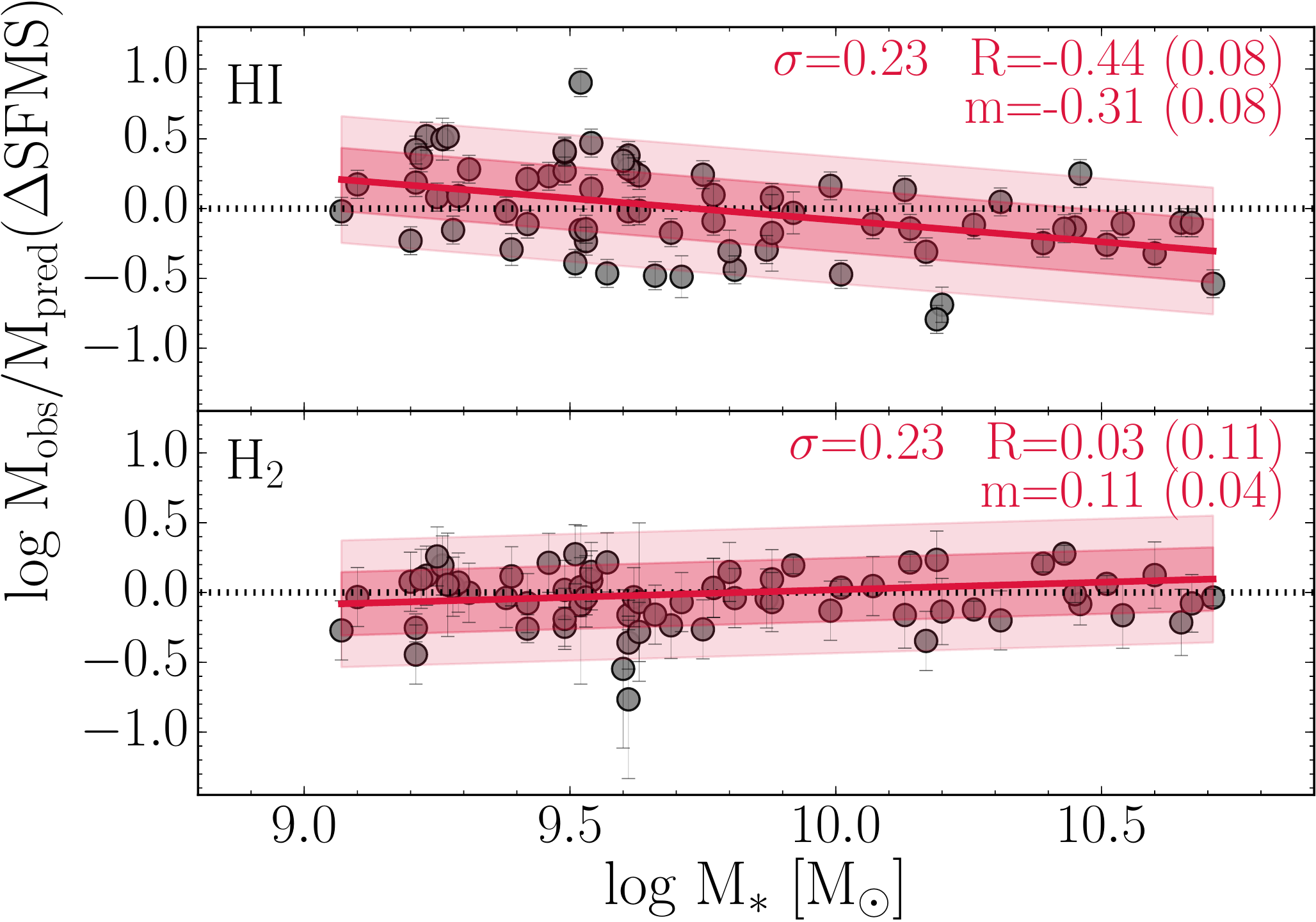}\\      
\vspace{5pt}
\includegraphics[width=7.75cm]{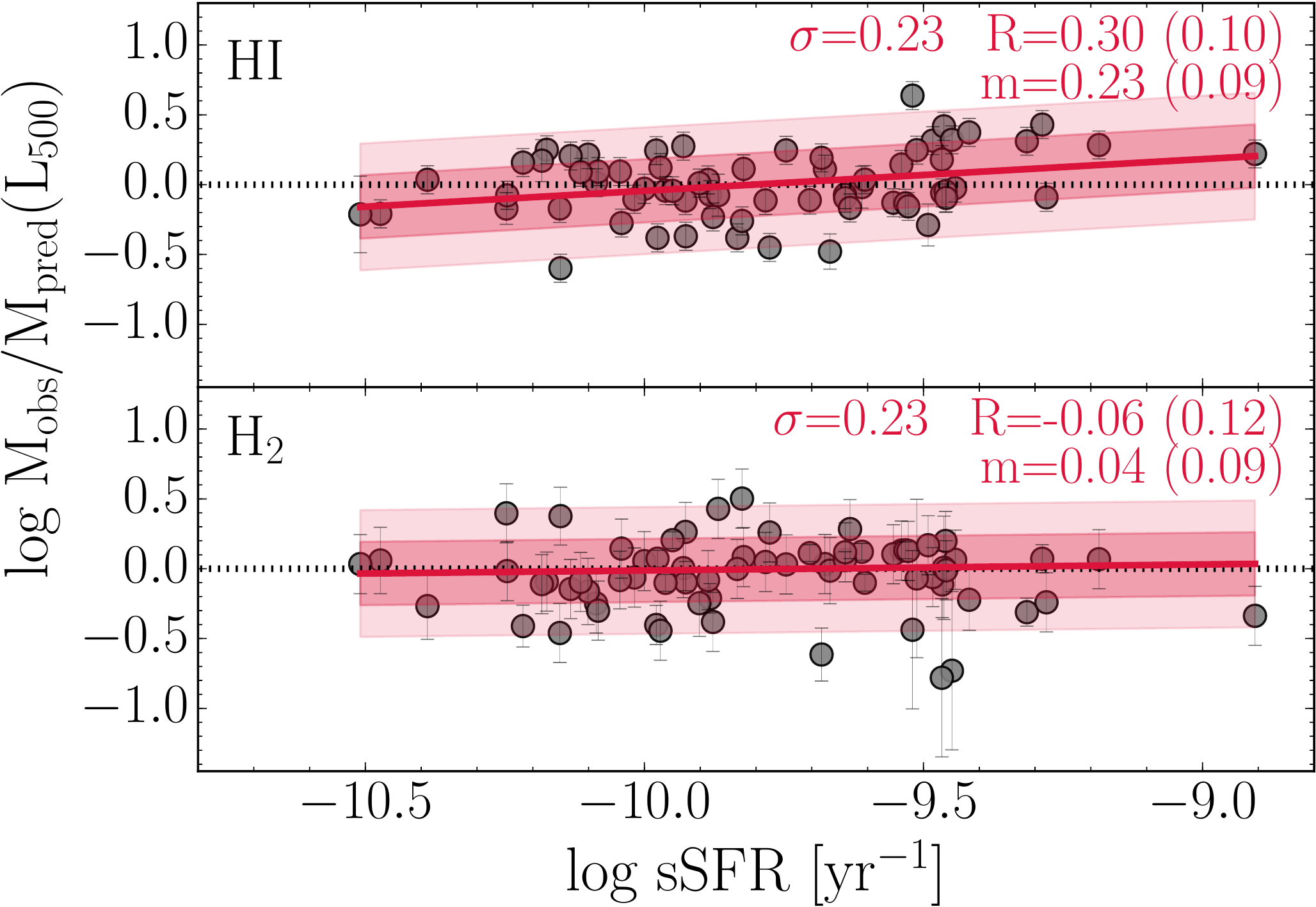} \hspace{15pt}  
\includegraphics[width=7.75cm]{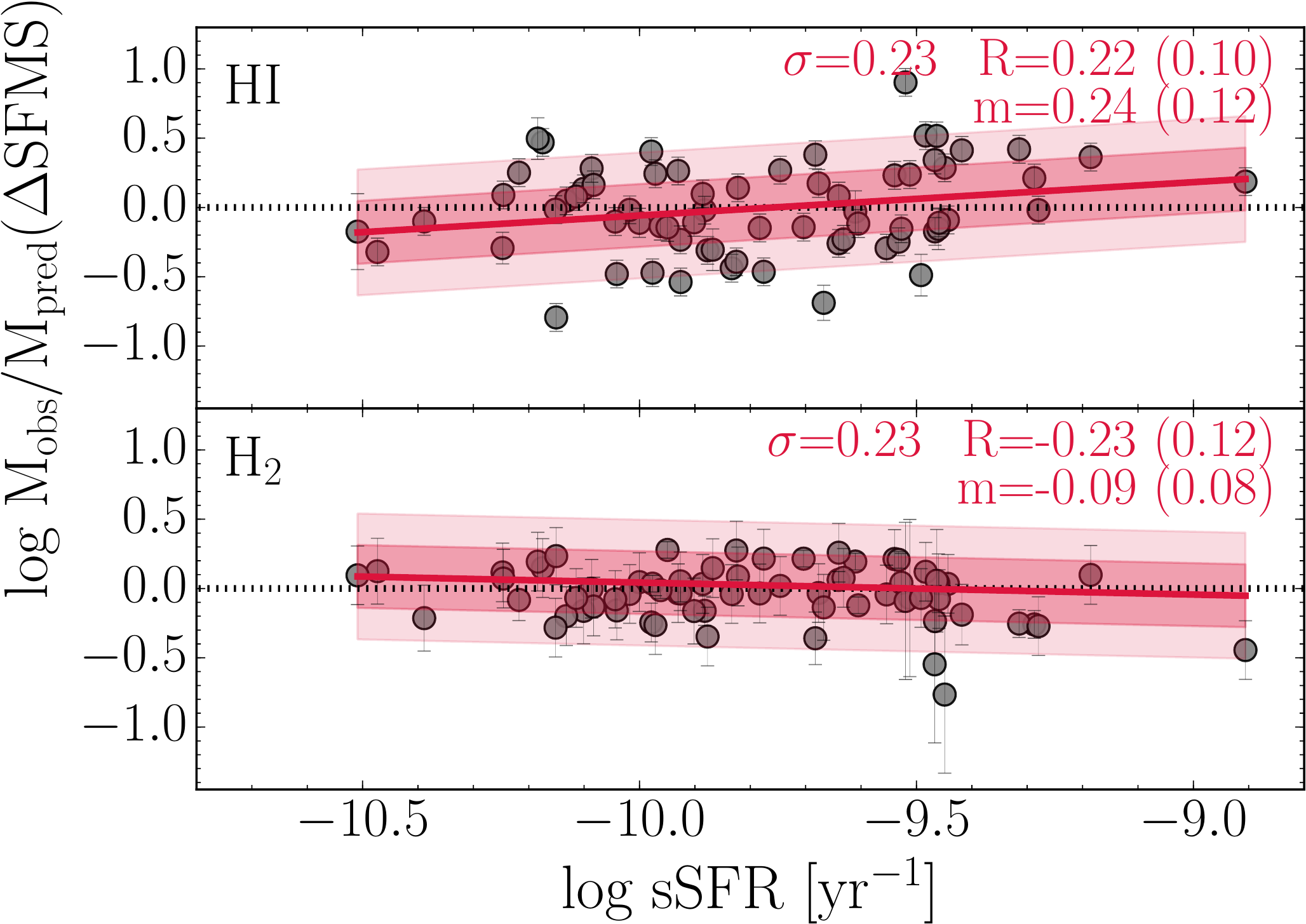}\\      
\caption{
Left panels: Residuals of \L500 (dust) predictions \added{for \mhi \,
  and 
  \mH2 as a function of the mass of the other phase. Top panel shows
  \mhi \, divided by 
    its prediction from \L500, demonstrating an anti-correlation
    with \mH2. Bottom panel shows \mH2 divided by its
    prediction from \L500, with similarly strong anti-correlation with
    \mhi. }
%
Right panels: Residuals of the $\Delta$SFMS (depletion time)
predictions for \added{\mhi \, and \mH2 as a function of the mass of
  the other phase. A stronger residual trend is found when predicting
  \mhi \, from depletion time, while no trend exists in the \mH2
  prediction.}
\label{fig:resid}
}
\end{figure*}

\subsection{Residuals between phases}

\added{
  %
  We quantify the underlying dependence between predictions in each
  phase by  plotting 
%
%
the accuracy of the predictive relationships for \mhi \, and \mH2 as
functions of \mH2 and \mhi, respectively. In this way we are comparing
observationally independent quantities, and the residual trends
can be meaningfully interpreted (see Appendix~\ref{sec:app}
for more discussion of this method).
The top two panels in the left column of
Figure~\ref{fig:resid} show the differences
between the \L500-based prediction and the real gas masses 
for \hi \, and \H2. The \mhi \, differences are plotted against
\mH2, and the \mH2 differences against \mhi \, (i.e., 
independently observed quantities). 
Each residual trend is fitted and
their slopes (m), scatters ($\sigma$), and correlations (R) are
shown, with uncertainties. These fits use least
squares minimization of the ordinate 
and are weighted by the uncertainties of the observed gas phase masses
only (i.e., they do not include the uncertainties on the predicted gas
masses).}

\added{
  For galaxies with larger \mH2, \L500
under-predicts \mhi, and for galaxies with larger \mhi, \L500
under-predicts \mH2. This behavior illustrates that \emph{\L500 is
  most tightly correlated with the total gas mass} (e.g., see
Section~\ref{sec:mgas}); any \mhi \, or \mH2 prediction based on
\L500 will have a systematic uncertainty depending on the partition of
atomic and molecular gas. 
These systematic trends are of modest amplitude (slopes of 17-18\% per
dex, with $\ge$2$\sigma$ significance) and large scatter ($0.2$~dex),
but are evident even within our small sample of local star-forming
galaxies. At higher redshift and for larger more star-forming systems,
extrapolations of these discrepancies could be larger.
}
%
%


\begin{figure*}
\includegraphics[width=0.99\textwidth]{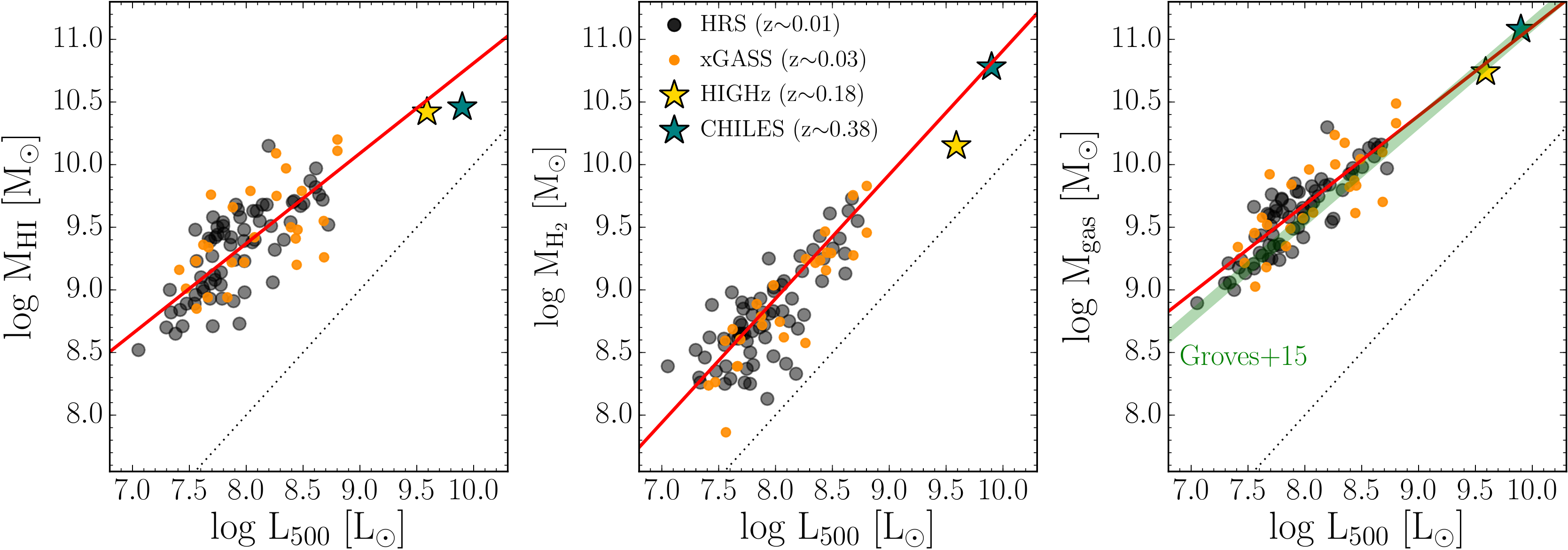} 
  \caption{
  All three panels show galaxies in our sample (in grey) and other
  colors show other
  galaxies including some 
  at higher redshifts. The dotted line shows the 1:1 relationship. Our
  best-fitting relationships for \mhi \, 
  (left), \mH2 (center), and \mgas \, (right) are shown as red
  lines. Higher redshift galaxies 
  appear to follow similar behavior as our local sample, with a tight
  \L500-\mgas \, relation and correlated scatter in the \mhi \, and
  \mH2 relations.
    \label{fig:highz}
  }
\end{figure*}

\added{In a similar way, the top two panels in the right column of
  Figure~\ref{fig:resid} 
  compare the depletion time-based predictions of gas masses (using
  $\Delta$SFMS) for \mhi \, and \mH2 with 
  their true values as a function of \mH2 and \mhi,
  respectively. The
discrepancies in \mhi \, predictions show steep systematic variations
(slope of 40\% per dex with $\ge4$$\sigma$ significance), which is not
unexpected since depletion time 
is less physically connected to the neutral atomic gas content.
Conversely, the errors in the \mH2 prediction do not depend on \mhi,
as this method is best suited to predict \mH2, the gas which is
directly involved in star formation. When using SFR to predict gas
mass, there 
is an implicit assumption about the rate of \H2 conversion into
stars as well as the conversion between \hi \, and \H2. This 
introduces an underlying dependence on the partition between
  \mhi \, and \mH2, as was seen in the M/L estimates.
}
%
%


These residual trends serve as a reminder that \added{ISM mass
  predictions from dust luminosities and star formation rates are not 
  equivalent and are sensitive to different gas phases.
In the worst cases,
using \L500 to predict \mH2 is dependent on \mhi \,
($R=-0.55$) with a modest slope ($m=0.17$), giving systematic 
over/under-predictions of $\sim$0.15~dex at the extremes of our
sample. Similarly, using 
t$_\textrm{dep}$ to predict \mhi \, results in an even steeper slope
($m=-0.40$) and similar correlation ($R=-0.48$), over/under-predicting
by up to $\sim$0.3~dex. Without prior
knowledge of the partition between molecular and atomic gas it is
difficult to apply these predictions.} Other studies have
shown that the \mH2/\mhi \, ratio in galaxies can depend on the
metallicity and turbulence of their ISM, which affect the conversion
between atomic and molecular gas \citep{krumholz09, bialy17}.

\subsection{Residuals with other physical properties}
\label{sec:phys}

\added{We also explored similar residual dependences 
  using other galaxy properties. In particular, we considered
  galaxy stellar mass 
  and sSFR. The lower four panels of 
  Figure~\ref{fig:resid} show, as a function of these physical
  properties, the ratios between the observations of
  \mhi \, and \mH2 and their \L500 and $\Delta$SFMS-based
  predictions. As with the top panels, we fit these residual trends to
  quantify their strengths.
}

\added{
  For both \L500 and $\Delta$SFMS, we see negative (positive)
  residuals for the \mhi \, (\mH2) predictions with increasing stellar
  mass. While sometimes weak, these trends are expected, given the
  moderately increasing \mH2/\mhi \, ratio observed as a function of
  stellar mass (Catinella \etal submitted).}

\added{All of the residual
  trends with \Mst \, are weaker (smaller $R$ values and almost
  always flatter slopes) than those with \mhi \, or \mH2, indicating
  that a stellar mass dependence is not enough to account for the
  residuals observed between the phases. The sSFR residuals in the
  predictions of \mhi \, suggest that elevated star formation results
  in \L500 and $\Delta$SFMS) under-predicting \mhi, while in more
  passive galaxies, it may be over-predicted (by $\sim$0.2~dex, even
  within our sample's relatively small range of sSFR).  }


\added{These modest systematic trends with \Mst \, and sSFR suggest that
  using \L500 or $\Delta$SFMS as ISM mass 
  estimators will be affected by underlying dependences on other
  galaxy properties. Our calibrations are naturally best-suited
  to predicting cold gas masses for galaxies which are similar to
  those included in our sample. Any extrapolation or extension of
  these relationships to significantly different populations of
  galaxies (e.g., with higher or lower \Mst \, or
  sSFR) may suffer from systematic biases.
  Nonetheless, these predictions are robust and can reliably generate
  indirect estimates of cold gas masses.  }

%
%
%
%

\section{Scientific implications}
\label{sec:implications}


\subsection{Applications to high-z galaxy observations}

 \added{Figure~\ref{fig:highz} shows the basic relations between gas masses
and \L500 for our sample and galaxies at other redshifts where all
three observations (21cm, CO, and 500$\mu$m) are available. In
addition to the sample used in this work, we show 26 galaxies from the 
xGASS sample ($z$=0.01-0.05, Catinella \etal submitted) which have
\L500 observations from the NASA/ IPAC Infrared Science
Archiv}e,\footnote{\url{http://irsa.ipac.caltech.edu/}}
\added{survey data from
\textit{Herschel}-ATLAS \citep{valiante16,bourne16} and the
\textit{Herschel} Stripe 82 
Survey \citep[HERS]{viero14}.
}

 \added{We also include two galaxies with  observations at higher
redshifts ($z$>0.1). First, AGC~191728 
comes from the HIGHz sample of \citet{highz} at
$z$=0.176. 
Its \hi \, observations come from Arecibo and its CO(1-2)
observations from the Atacama 
Large Millimeter Array  \citep{cortese17}. This galaxy was
serendipitously imaged 
as part of the \textit{Herschel}-ATLAS observations 
and released in H-ATLAS DR1 \citep{valiante16}. 
}

\added{
Second, COSMOS~J100054.83+023126.2 comes from the Cosmic Evolution
Survey \citep[COSMOS][]{scoville07} and is at $z$=0.376 (this is the
highest redshift detection of \hi \, emission from a galaxy to
date). It has 
recently been observed in 21cm with the Jansky Very Large Array as
part of the COSMOS \hi \, Large Extragalactic Survey
(CHILES) 
 and in CO with the Large Millimeter Telescope \citep{fernandez16}.
In another serendipitous observation, DR2 of the \textit{Herschel}
Multi-tiered Extragalactic Survey \citep[HERMES,][]{oliver12} includes
500$\mu$m observations of this galaxy, with  fluxes available
through their SUSSEXtractor catalog.
}

\added{Note that the observed 500$\mu$m emission from these two
  sources corresponds to rest-frame observations at 425$\mu$m and
  363$\mu$m, respectively. While we use these fluxes in our
  \L500-based relationships, adopting the L$_{350}$-based predictions
  would make only a small difference. Within our sample, the ratio of
  L$_{350}$/\L500 is 1.073$\pm$0.006.
  }


\added{
These two higher redshift galaxies provide an illustrative example of
the application of the \L500-based predictions of cold gas
masses. These two galaxies are quite different: while they have very
similar \mhi \, (and stellar mass estimates from optical photometry),
the CHILES galaxy has $\sim$4 times larger \mH2 than the HIGHz
galaxy. Their \mH2/\mhi \, ratios are 50\% and 200\%,
respectively. Remarkably, the \L500 prediction for \mgas \, is very
accurate for both! However, \L500-based estimates of \mhi \, and \mH2
can be wrong by $\sim$0.4~dex. The \hi-dominated HIGHz galaxy lies
along the \L500-\mhi \, relationship, but falls $\sim$0.4~dex below
\L500-\mH2. Conversely, the \H2-dominated CHILES galaxy is consistent
with the \L500-\mH2 relation but falls $\sim$0.3~dex below
\L500-\mhi. \emph{Without additional knowledge of the molecular-to-atomic
  ratio of these galaxies, \L500 cannot reliably predict their \mhi \,
  or \mH2 separately.}}

\added{Extending this argument further,} observations have suggested
that  
higher-redshift galaxies have higher \mH2/\mhi \, ratios compared with
the \hi-dominated galaxies observed at low-redshift, but within
$z<0.2$ this ratio seems to \added{span the same range of values}
\citep{cortese17,cybulski16}. When using 
these types of relations to predict molecular gas masses at high
redshift, corrections may be necessary to account for different mass
contributions from ISM phases.
%
%
When predicting gas masses in high redshift
galaxies, appropriately calibrated relations must be used. While it is
possible to separate the atomic and molecular gas phases in our local
sample, the distinction becomes difficult at higher redshift. There is
inherent uncertainty in extending a low-redshift prediction to higher
redshift galaxies, which may have \added{different atomic-to-molecular
  ratios or other physical differences from local galaxies.}

\begin{figure}
\includegraphics[width=0.999\columnwidth]{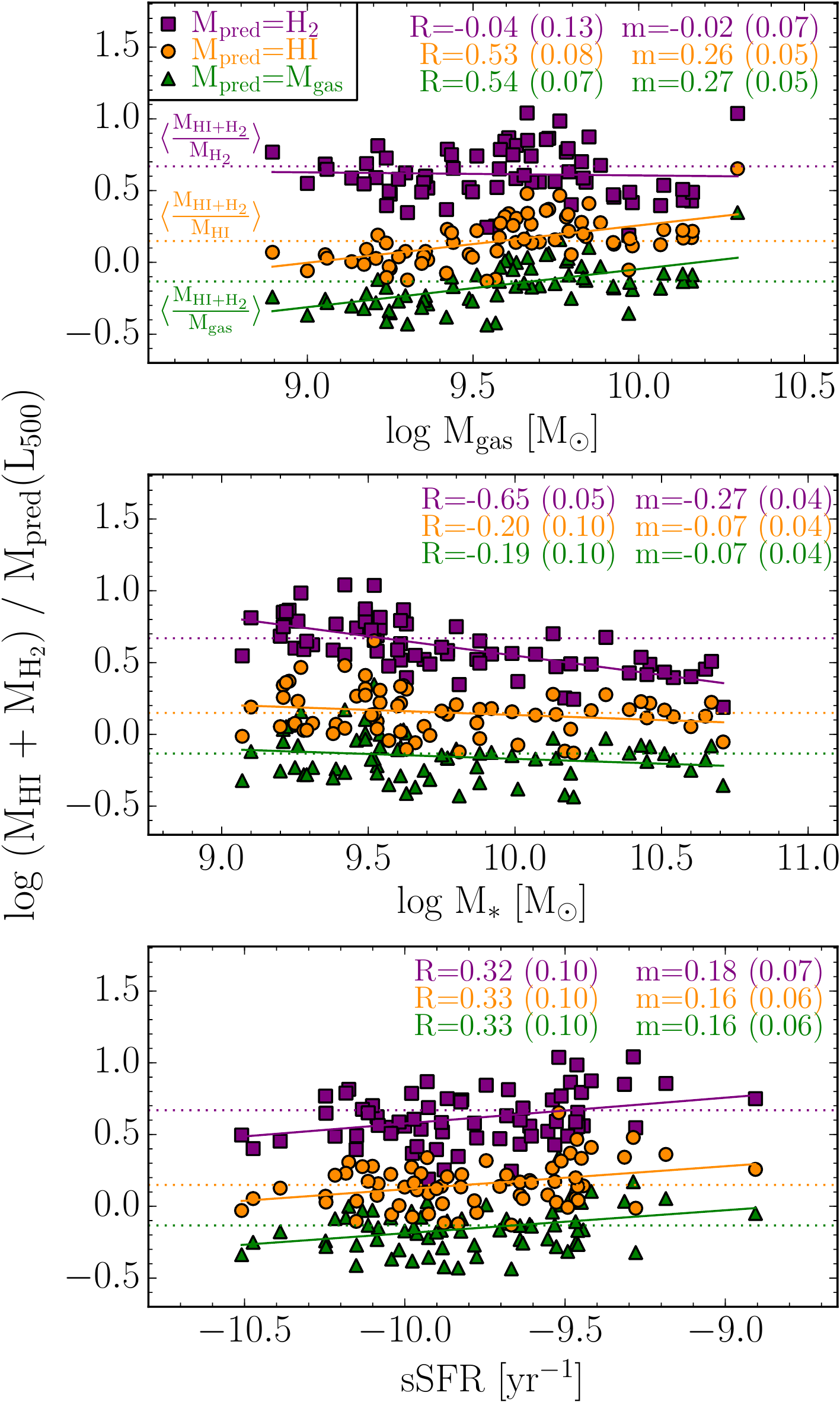} 
  \caption{
 Each panel plots the ratio between the observed \mhi+\mH2 and the
 \L500-based predictions of \mhi \, (\mH2) as orange circles (purple
 squares), as functions of total gas mass, stellar mass, and
 sSFR. Relations with \mgas \, are shown as green triangles. Least
 squares fits are shown along with best-fit slopes and
 uncertainties.
    \label{fig:gas2}
  }
\end{figure}

\subsection{Consequences of incorrect phase assumptions}

\added{In Figure~\ref{fig:gas2} we quantify the possible errors
  introduced by 
  applying \L500-based gas mass predictions to galaxies
  without knowledge of their molecular-to-atomic partitions. In each
  panel we either apply the assumption that galaxies are
  \hi-dominated and use the relationship calibrated for \mhi \,
  (orange circles) or are \H2-dominated and use the \mH2 
  relationship (purple  
  squares). For completeness we also show the comparison between
  total gas mass predictions and the observed \mhi+\mH2 (green
  triangles). Each of these comparisons has a systematic offset from
  unity: the \mH2 and \mhi \, predictions scatter around the average
  molecular and atomic gas fractions for this sample, and the \mgas \,
  predictions include the $\sim$30\% correction for the contribution
  of helium.
}

\added{
  When adopting the \hi-dominated assumption (orange circles), the
  y-axis shows the  
  difference between the true \mhi+\mH2 and the \L500-predicted
  \mhi, as \mH2 is assumed to be negligible.
  This assumption is most accurate for galaxies with
  \mgas$\sim$10$^9$\msun, with large stellar mass, or with passive
  sSFR, where the \L500-\mhi \, relationship accurately predicts the
  total gas mass. Estimates generated from this assumption become
  significantly 
  worse for more star-forming galaxies or those with larger gas
  masses (i.e., similar to those observed at higher
  redshifts) where the predictions can be up to $\sim$0.5~dex too
  small. For galaxies with more extreme \mH2/\mhi \, ratios, the
  effects of this incorrect assumption could be even larger.
}

\added{
  While the \H2-dominated assumption (purple squares) is clearly not
  valid for any of the galaxies in our sample (note that none reach
  unity in the ratio being plotted),  this assumption interestingly
  shows strong  systematic variations with \Mst \, and sSFR. These
  systematic trends   are stronger and more significant than those
  found in Figure~\ref{fig:resid}, and are a manifestation of the
  incorrect assumption about the dominant phase.
  Even the \L500-\mgas \, relationship (green triangles) shows a
  dependence on sSFR at the $\sim$2.5$\sigma$ level, demonstrating
  the ubiquity of underlying systematic residuals.
}

\added{
  While it is not unsurprising that incorrect assumptions about the
  dominant gas phase yield incorrect
  results, these trends demonstrate quantitatively the nature of
  errors arising from these assumptions. Most importantly, in addition
  to the expected systematic offset, Figure~\ref{fig:gas2} shows that
  residual trends are present which 
  depend on other galaxy properties. \emph{Every
    application of these indirect gas mass predictions relies on
    implicit assumptions to produce \mhi \, or \mH2 estimates.} 
}
%
%
 %
%


\subsection{The K-S with different calibrations}

\begin{figure}
\includegraphics[width=0.98\columnwidth]{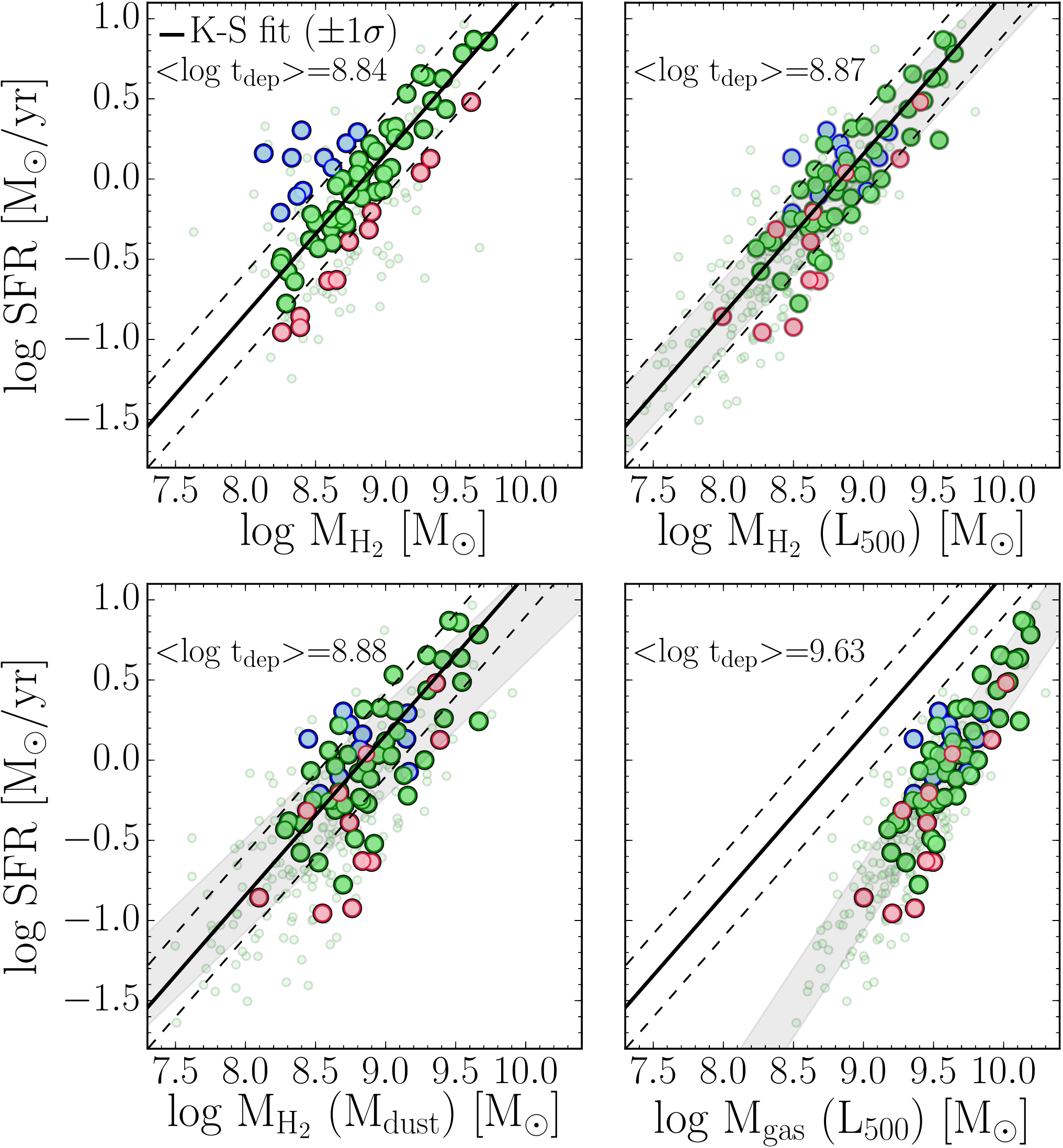}
  \caption{Integrated star formation law with different calibrations. Top left
    panel shows the true relation between the observed \mH2 and SFR,
    with the best-fit relationship (shown in all panels). 
       Other panels plot observed SFR against various 
    estimates of ISM phase masses, with \added{analogous} best-fits
    shown as grey-shaded 
    regions. Galaxies 
    are color-coded based on their position on, above, or below the
    relation. The full sample of HRS galaxies is included as faint
    points whenever available.
    \label{fig:ks}
  }
\end{figure}

Figure~\ref{fig:ks} shows the
integrated K-S star formation law \citep{schmidt59,kennicutt83},
using different predictors of ISM  
phase massses. While the top left panel shows the K-S law from direct
observations of SFR and \mH2, the other panels use different indirect
gas mass estimates.

Note that the slope, width, and offset of the
best-fit relation can all be significantly affected by a different
choice of gas mass indicator. For example, when adopting an estimate
of \mH2 based on \L500, the K-S relation becomes steeper and tighter
than the true relation. 
Also note the extent to which an individual galaxy can move
above/below the K-S relation using different gas 
estimates. The different-colored points on Figure~\ref{fig:ks}
demonstrate that >$+1\sigma$ outliers can become consistent with the
relation (using \mH2 based on \L500) and that <$-1\sigma$ outliers can
move above the relation (using \mH2 based on \mdust). 

\added{The lower right panel of Figure~\ref{fig:ks} shows what can
  happen when using total \mgas \, instead of the star-forming \mH2. This
illustrates the impact of an incorrect assumption that the ISM is
dominated by the molecular component when in fact it is mostly
atomic. While this assumption is obviously inappropriate for the local
\hi-dominated galaxies in our sample, higher redshift galaxies have
a wide range of \mH2/\mhi \, partitions and may not all be dominated
by the molecular component \citep{cortese17}.}

It can be dangerous to use these types of indirect estimates of gas
masses to study the K-S relationship or other scaling
relations. \emph{Any calibration of these gas mass  
estimates will be linked to the underlying physical properties of the ISM}
(e.g., atomic-to-molecular hydrogen ratio), and will not necessarily 
indicate variations in star formation efficiency.
There are likely to be systematic effects lurking 
in these calibrations which will limit any attempt to derive a K-S
relation using indirect gas mass predictions.

\section{Summary}
\label{sec:summary}

Using a representative sample of N=68 nearby galaxies from the
\textit{Herschel} 
Reference Survey, we have calibrated a set of relationships
between the masses of ISM phases and observable quantities (FIR and
SFR). These predictive relationships can estimate ISM masses with
$\sim$20\% accuracy and are mutually self-consistent. However, our
complete set of observations of all ISM phases \added{show} that these 
predictive relationships suffer from modest \emph{systematic residual
  dependences \added{on the molecular-to-atomic partition and other
    physical 
  properties}}. Any 
application of these relationships to 
predict gas masses from FIR/SFR observations requires
an implicit assumption of the underlying \mH2/\mhi \,
ratio. \added{Incorrect assumptions about the dominant phase of the
  ISM can yield errors in gas mass predictions as large as 
  0.5~dex.} Furthermore,
using these indirect gas estimates to test the evolution of
star formation laws or other scaling relations is potentially
problematic, as these relations rely on those underlying scaling
relations to successfully predict gas masses.

\section*{Acknowledgements}


We thank Toby Brown and Katinka Ger\'eb for helpful discussions, and
the anonymous referee for their comments which have significantly
improved this work and its statistical treatment of residual
trends.

SJ, BC, and LC acknowledge support from the Australian Research
Council's Discovery Project funding scheme (DP150101734).
BC is the recipient of an Australian Research Council Future
Fellowship (FT120100660). AG acknowledges support from the ICRAR
Summer Studentship Programme during which this project was initiated.

This research has made use of NASA's Astrophysics Data System, and
also the 
NASA/IPAC Extragalactic Database (NED), which is operated by the Jet
Propulsion Laboratory, California Institute of Technology, under
contract with the National Aeronautics and Space Administration. 
This research has also made use of the NASA/ IPAC Infrared Science
Archive, 
which is operated by the Jet Propulsion Laboratory, California
Institute of Technology, under contract with the National Aeronautics
and Space Administration.
Finally, this
research has made extensive use of the invaluable Tool for
OPerations on  Catalogues And Tables
\citep[TOPCAT\footnote{\url{http://www.starlink.ac.uk/topcat/}}, 
][]{taylor05}.





%



\appendix

\section{Caveats to detecting systematic trends}
\label{sec:app}

We undertook a Monte Carlo (MC) analysis to more fully understand and
interpret the nature of the 
residual systematic trends explored in this work. One of our main
points is the importance of the \mH2/\mhi \, ratio in applying
indirect predictions of cold gas masses. However, this MC analysis
demonstrates that one must be cautious when exploring dependences on
that ratio, as certain techniques will show apparent residual trends
which are partially specious and misleading. To verify that our
results are not affected by non-physical trends, we describe our MC
approach and analysis below.

\begin{figure*}
\centering
\includegraphics[width=0.95\textwidth]{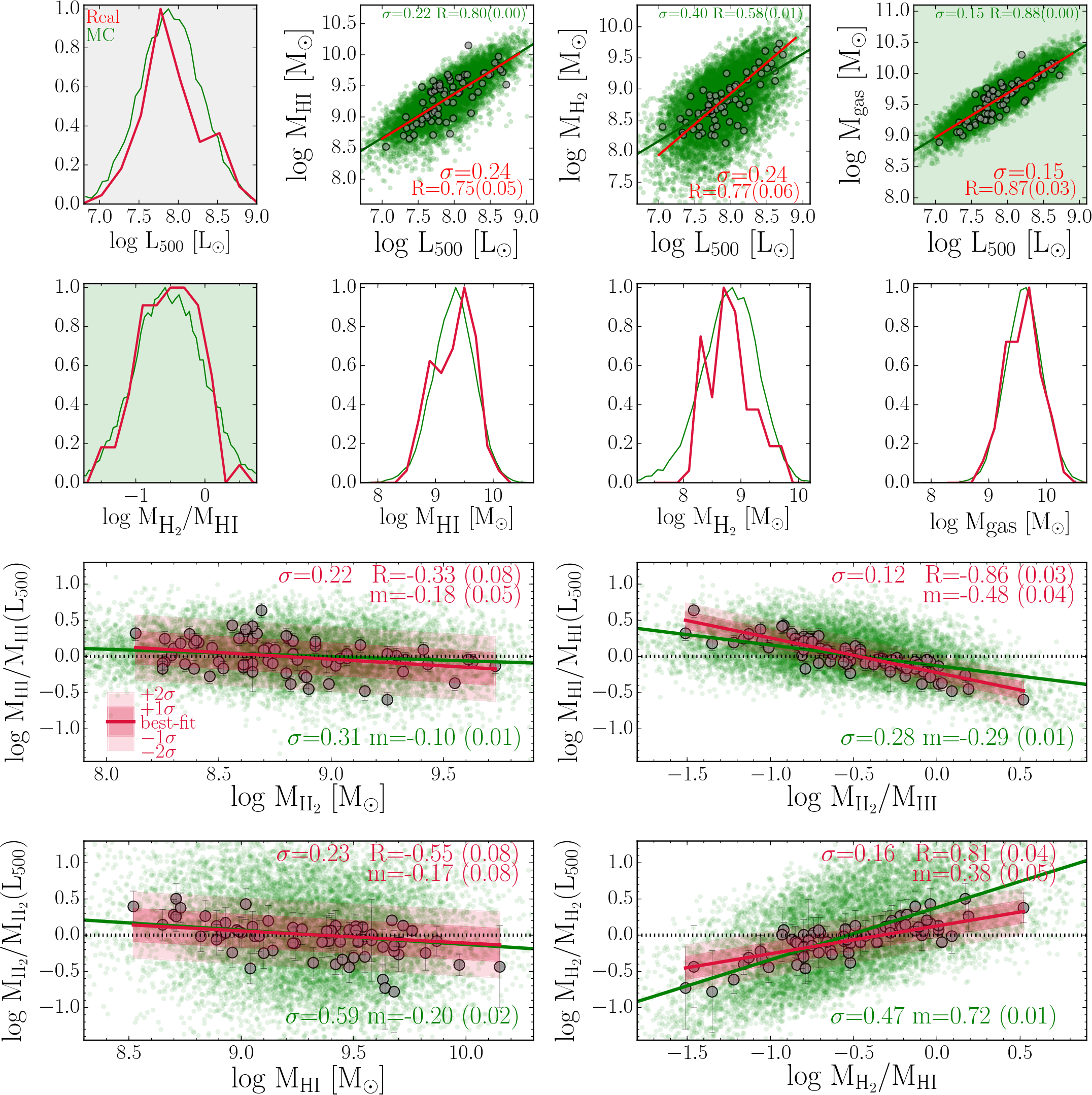} 
  \caption{Full summary plots of our Monte Carlo for \L500-based
       predictions of cold gas masses. In all panels, grey points and
       red lines show data, histograms, and fits to our real
       observations; green points and lines show the simulated data.
    \label{fig:app1}
  }
\end{figure*}


To simulate and test the relationships between \L500 and cold gas
masses, we start by generating a random distribution of \L500 values
(N=10,000) 
consistent with that of our sample. We use our relationship between
\L500 and \mgas \, to generate estimates for the total gas content,
and add random noise to the estimates to match our observed dispersion
of $\sigma$=0.15~dex. We next generate a distribution of \mH2/\mhi \,
values which matches that of our sample, ranging from $\sim$3\% to
$\sim$300\%, and calculate \mH2 and \mhi \, for each value of \mgas \, 
using this ratio. The top two rows of Figure~\ref{fig:app1} show
histograms of these quantities and the relationships with \L500, where
grey points and red lines show our observations and green points and
lines show the simulated MC data. Note
that we only input the \L500-\mgas \, relation into this MC, and we
naturally recover the observed slope, correlation strength, and
scatter in the
\L500-\mhi \, relation. However, the resulting MC version of the
\L500-\mH2 relation does not agree as well with the observed relation,
although the two distributions of points have significant overlap.

The bottom two rows of Figure~\ref{fig:app1} show two different
techniques (i.e., quantities on the x-axes) for
quantifying residual trends in our \L500-based predictions of \mhi \,
and \mH2. The left two panels show the same type of analysis as used
in this work: the ratio of \mhi \, to its prediction is plotted
against \mH2, and vice versa for the \mH2 prediction ratio against
\mhi. In both cases, the slope of the MC residuals is similar to our
observations, although our observed trend with \mhi/\mhi(\L500) is
slightly larger than the MC value and the scatter about the
\mH2/\mH2(\L500) residual is over twice as large as our observations.

The right two panels show a different approach to measure the
residuals as a function of \mH2/\mhi, which is expected to drive these
trends. However, since \mhi \, (or \mH2) appears on both axes, both
the real and MC sample show enhanced trends as a function of
\mH2/\mhi. While not physical, this type of plot is useful when
interpreting \L500-based predictions for galaxies with different
molecular-to-atomic gas mass ratios. Nonetheless, \textit{caution is
  advised when quantifying residual trends in \mhi \, predictions as a
  function of \mH2/\mhi.}

The agreement of this MC analysis with our observations suggests that
the strongest predictive 
power comes from the \L500-\mgas \, relationship. This MC
analysis goes one step further and shows that any attempt to use \L500
to predict \mhi \, or \mH2 alone will suffer from a systematic dependence
on the (potentially unknown) \mH2/\mhi \, ratio. We also note that
this MC analysis does not produce results which fully agree with our
observations, indicating that further residual dependences on other
physical properties (e.g., those discussed in Section~\ref{sec:phys})
may also play a role in these relationships.


\bsp	
\label{lastpage}
\end{document}